\documentclass[twocolumn]{aastex631}

\usepackage{times}
\usepackage{amsmath,amssymb,amstext}
\hypersetup{linkcolor=red,citecolor=blue,filecolor=cyan,urlcolor=blue}
\usepackage{tabularx}
\usepackage{longtable}
\usepackage{float}
\usepackage{natbib}
\usepackage{graphicx,color}
\usepackage{txfonts}

\usepackage{multirow}
\usepackage{array}
\usepackage{rotating}
\usepackage{sidecap}
\usepackage{hyperref}
\usepackage{epstopdf}
\usepackage{footnote}
\usepackage{tabularx}
\usepackage{booktabs}
\usepackage{longtable}
\usepackage{tabu}
\usepackage{longtable}

\newcommand{\kms}{km\,s$^{-1}$}

\begin{document}
\title{{\large {X-ray/Radio Quasi-periodic Pulsations Associated with Plasmoids in Solar Flare Current Sheets}}}
\author{Pankaj Kumar\altaffiliation{1,2}}

\affiliation{Department of Physics, American University, Washington, DC 20016, USA}
\affiliation{Heliophysics Science Division, NASA Goddard Space Flight Center, Greenbelt, MD, 20771, USA}

\author{Judith T.\ Karpen}
\affiliation{Heliophysics Science Division, NASA Goddard Space Flight Center, Greenbelt, MD, 20771, USA}

\author{Joel T.\ Dahlin}
\affiliation{Heliophysics Science Division, NASA Goddard Space Flight Center, Greenbelt, MD, 20771, USA}
\affiliation{Astronomy Department, University of Maryland, College Park, MD 20742, USA}
\email{pankaj.kumar@nasa.gov}

\begin{abstract}
Plasmoids (or magnetic islands) are believed to play an important role in the onset of fast magnetic reconnection and particle acceleration during solar flares and eruptions. Direct imaging of flare current sheets and formation/ejection of multiple plasmoids in extreme ultraviolet (EUV) images, along with simultaneous X-ray and radio observations, offers significant insights into the mechanisms driving particle acceleration in solar flares. Here we present direct imaging of the formation and ejection of multiple plasmoids in flare plasma/current sheets and associated quasi-periodic pulsations (QPPs) observed in X-ray and radio wavelengths, using observations from SDO/AIA, RHESSI, and Fermi GBM. 
These plasmoids propagate bidirectionally upward and downward along the flare current sheet beneath the erupting flux rope during two successive flares associated with confined/failed eruptions. The flux rope exhibits evidence of helical kink instability with formation and ejection of multiple plasmoids in the flare current sheet, as predicted in an MHD simulation of a kink-unstable flux rope.  RHESSI X-ray images show double coronal sources (``loop-top" and higher coronal sources) located at both ends of the flare current/plasma sheet. Moreover, we detected an additional transient faint X-ray source (6-12 keV) located between the double coronal sources, which was co-spatial with multiple plasmoids in the flare current sheet. X-ray (soft and hard) and radio (decimetric) observations unveil QPPs (periods$\approx$10-s and 100-s) associated with the ejection and coalescence of plasmoids. These observations suggest that energetic electrons are accelerated during the ejection and coalescence of multiple plasmoids in the flare current sheet. 
\end{abstract}


\section{INTRODUCTION}\label{intro}
 Magnetic reconnection is the fundamental process for energy release in solar flares and coronal mass ejections \citep{shibata2011,chen2011}.  During reconnection, the magnetic energy stored in sheared magnetic fields is rapidly released in the form of thermal energy, kinetic energy, and particle acceleration \citep{benz2017}.
 Traditional models of reconnection, such as the Sweet-Parker model, predict reconnection rates that are too slow to account for the rapid energy release observed in solar flares. The Petschek reconnection model \citep{Petschek1964} predicts fast magnetic reconnection via slow-mode shocks, but it requires the formation of a very specific and narrow diffusion region, which is difficult to sustain over large scales or in resistive MHD conditions. To resolve this discrepancy, more recent models have introduced the concept of plasmoids that form within the current sheet due to the tearing mode instability. As the current sheet thins and stretches prior to reconnection, it becomes increasingly susceptible to this instability, leading to the formation of multiple plasmoids \citep{shibata2001, Loureiro2007, Bhattacharjee2009,huang2010,barta2011}. These plasmoids enhance the reconnection process by increasing the local reconnection rate, effectively speeding up the overall process and facilitating the rapid energy conversion needed to power solar flares. The presence of plasmoids also helps explain the observed fine structures in solar flare emissions and the quasi-periodic pulsations often detected in EUV, X-ray, and radio observations. These plasmoids play a crucial role at the high Lundquist numbers characteristic of flares, as their formation, motion, and eventual coalescence are closely linked to the acceleration of particles to high energies. Observing the emissions from these accelerated particles provides valuable insights into the underlying mechanisms of flare energy release. The acceleration of energetic electrons resulting from the ejection and coalescence of plasmoids in the current sheet has broad implications for understanding fast magnetic reconnection in solar, heliospheric, and magnetospheric current sheets \citep{chen2008,phan2024}.

Electron acceleration in the flare current sheet, particularly from plasmoids formed by reconnection, involves several key mechanisms. These electrons gain energy from the reconnection electric field, and from betatron and Fermi acceleration mechanisms within contracting plasmoids. Additionally, the turbulent environment created by interacting and merging plasmoids enhances stochastic acceleration processes \citep{drake2006}. Numerical simulations and observations continue to provide critical insights into the complex dynamics of electron acceleration in flare current sheets \citep{dahlin2014,dahlin2017, chen2020}. Particle-in-cell (PIC) and {\it kglobal} simulations suggest that the coalescence of multiple plasmoids can produce suprathermal particles \citep{oka2010,drake2013,arnold2021}. Alternatively, termination shocks may form when fast reconnection outflow collides with the top of the flare arcade, creating a site for efficient particle acceleration. These shocks accelerate electrons and ions through processes such as diffusive shock acceleration, where particles gain energy by repeatedly crossing the shock front \citep{aurass2002,aurass2004,chen2015}. This mechanism also can significantly contribute to the high-energy particle populations observed in solar flares.

Hard X-ray (HXR) emission in solar flares is primarily produced by high-energy electron beams that are accelerated during magnetic reconnection and interact with the dense low coronal and chromospheric plasma \citep{krucker2008}. Yohkoh and RHESSI (Reuven Ramaty High Energy Solar Spectroscopic Imager) observations have revealed the presence of loop-top (above the flare arcade) and footpoint sources during flare impulsive energy release \citep{masuda1994,dennis2022}. 
 RHESSI observations discovered the presence of coronal double X-ray sources during a few solar flare events \citep{sui2003,sui2004}, which enhances our understanding of the fundamental processes about energy transfer and particle acceleration in flares. This double source configuration was interpreted as evidence of magnetic reconnection in the flare current sheet, where the lower source is typically associated with the loop top source above the flare arcade and the upper source with the reconnection outflow high in the corona \citep{wang2007,liu2008}. The flare energy-release site is believed to lie between these double X-ray sources. The nature of coronal double X-ray sources is not yet fully understood. The simultaneous observations of flares in EUV, X-ray, and radio wavelengths offers a valuable opportunity to explore the mechanisms of energy release and particle acceleration. 

Quasi-periodic pulsations (QPPs) are a common and intriguing feature observed in flare emission across a wide range of wavelengths, including radio, X-ray, and extreme ultraviolet (EUV) \citep{kane1983,inglis2008,kumar2016,kumar2017a,hayes2020}. These pulsations typically manifest as periodic or quasi-periodic variations in intensity, with periods ranging from a fraction of a second to several minutes. The physical mechanisms driving QPPs are not well understood, but they are generally thought to be related to repetitive magnetic reconnection processes associated with magnetohydrodynamic (MHD) oscillations or bursty reconnection/plasmoids within the flaring region \citep{nakariakov2009,mcLaughlin2018,zimovets2021}. MHD simulations have demonstrated the formation and ejection of a series of plasmoids moving bidirectionally during explosive flare reconnection in a current sheet beneath erupting flux ropes \citep{barta2008,lynch2013,karpen2012, guidoni2016, dahlin2022}.

Decimetric radio bursts, typically observed in the frequency range of 300-3000 MHz, are produced by non-thermal electrons accelerated during the magnetic reconnection process \citep{benz2011}. The height of decimetric (236-432 MHz) radio sources observed by the Nançay Radioheliograph (NRH) generally lies above the RHESSI loop-top source \citep{vilmer2002,pick2005,benz2011}. 
Previous studies have proposed that decimetric radio bursts and pulsations (drifting towards lower frequencies) can be produced by the ejection and coalescence of multiple plasmoids in the flare current sheet \citep{kliem2000}. Drifting pulsating structures (DPSs) in decimetric radio bursts likely indicate the intermittent nature of energy release in the flare current sheet, and signature of radio emission from electrons accelerated during the ejection and coalescence of plasmoids \citep{karlicky2004,karlicky2010,karlicky2010a,karlicky2011}. 

A few previous studies have linked the behavior of plasmoids with the acceleration of electrons, leading to radio and X-ray emission, during magnetic reconnection in flares. 
RHESSI observations suggest an increase in hard X-ray and radio emissions during the coalescence of a downward-moving coronal X-ray source (interpreted as a plasmoid) with a looptop kernel \citep{milligan2010}. However, EUV images of this event do not reveal any plasmoids.
\citet{kumar2013} reported the first simultaneous EUV and radio (DPS in the decimetric band) observations of bidirectional plasmoids during an X-class flare. The speeds of upward/downward moving plasmoids observed in EUV and in radio DPSs (both positive and negative) were found to be consistent. \citet{takasao2016} presented radio imaging of microwave bursts (i.e., gyrosynchrotron emission) during the ejection and coalescence of downward-moving plasmoids along with their interaction with the flare arcade.

 This paper presents EUV imaging of plasmoids formed in flare current sheets underneath erupting flux ropes during two successive flares on 2015 April 22. The bidirectional plasmoids were associated with quasiperiodic pulsations in X-ray (soft/hard) and radio (decimetric) wavelengths. The flux rope appeared only in the hot channels (131/94 \AA) during the first flare. We present direct imaging of the formation of double coronal X-ray sources at both ends of the flare current sheet during the ejection and coalescence of multiple plasmoids. The erupting flux rope during the second flare apparently undergoes kink instability and the formation of a plasma/current sheet along with multiple plasmoids propagating bidirectionally. In both cases the flux ropes reached a height of about 45 Mm (60 arcsecs) above the limb but remained confined within the overlying strapping field of the active region, thus failing to produce CMEs.  In \S 2, we present the observations and results, while in \S 3 we discuss and summarize the results.

\begin{figure*}
\centering{
\includegraphics[width=8.4cm]{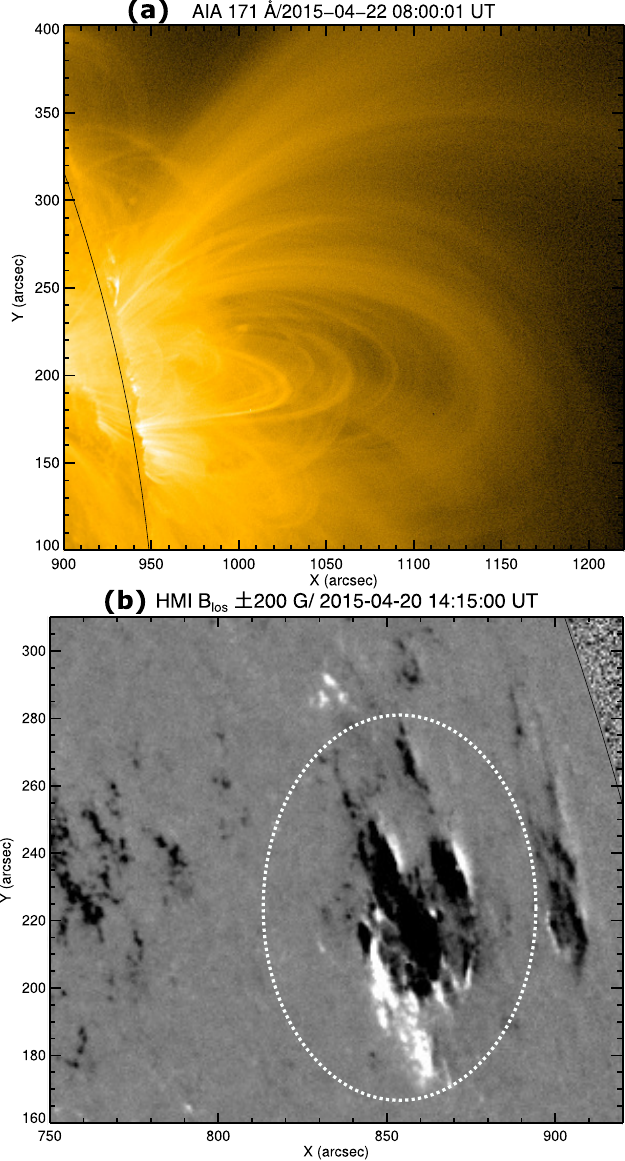}
\includegraphics[width=9.2cm]{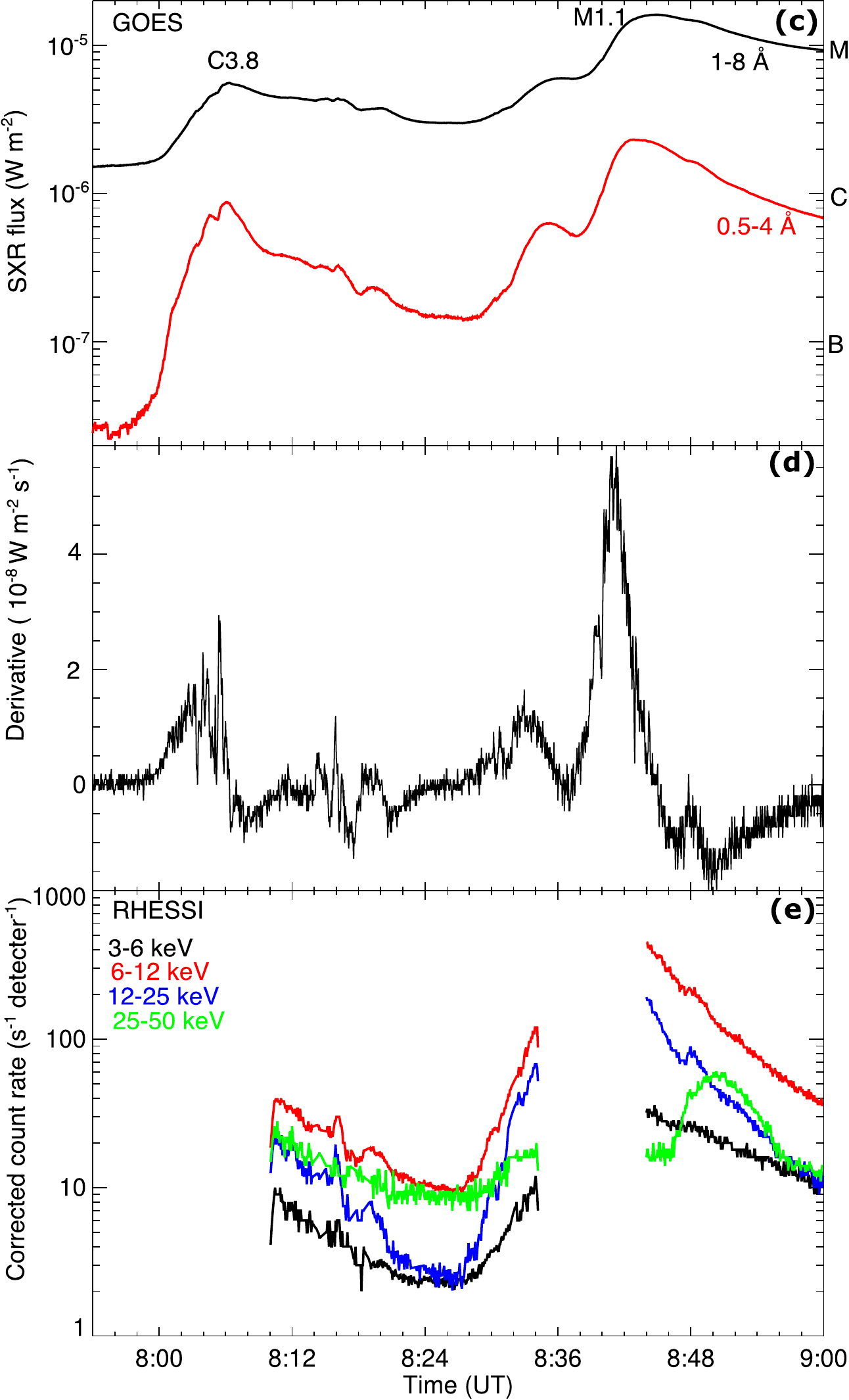}
}
\caption{{\bf NOAA active region 12322 and associated flares.} (a) SDO AIA 171 {\AA} image showing magnetic structures within the active region prior to the first eruption on 22 April 2015. (b) HMI magnetogram (line-of-sight photospheric magnetic field, scaled between $\pm$200 G) of the active region (inside white oval) two days before the flares (20 April 2015). (c) GOES soft X-ray flux profiles (2-s cadence) in 1-8 {\AA} (black) and 0.5-4 {\AA} (red) channels. (d) GOES soft X-ray flux derivative in the 1-8 {\AA} channel. (e) RHESSI X-ray flux profiles (4-s cadence) in 3-6, 6-12, 12-25, and 25-50 keV energy bands.} 
\label{fig1}
\end{figure*}
\begin{figure*}
\centering{
\includegraphics[width=18cm]{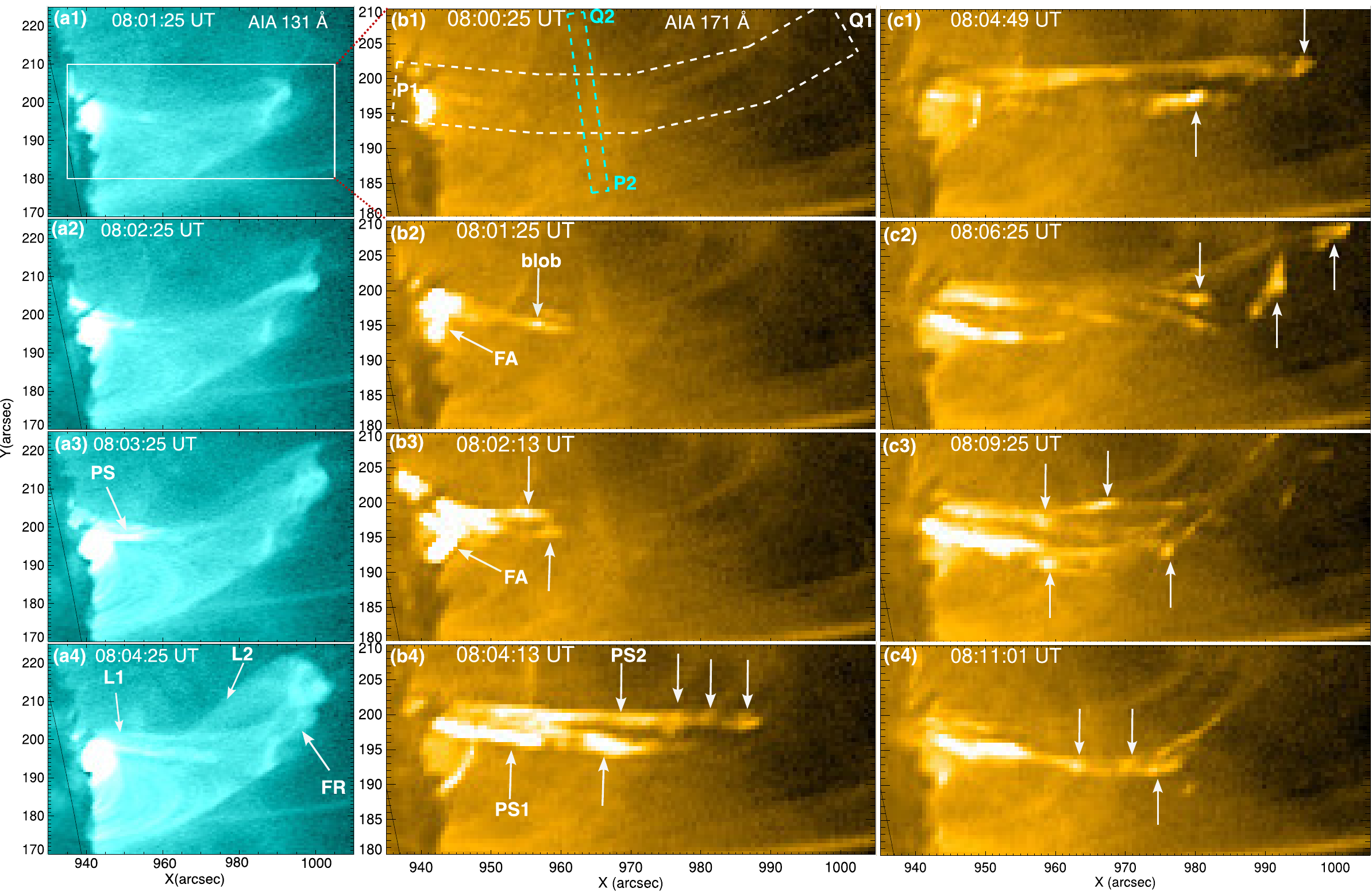}
}
\caption{{\bf Flux rope and blobs during the first flare (C3.8).} (a1-a4) AIA 131~{\AA} images showing the flare arcade (FA) and a flare plasma sheet (PS) below the hot flux rope (FR). L1 and L2 indicate the legs of the flux rope. (b1-b4,c1-c4) Sequence of AIA 171~{\AA} images showing the ejection of multiple blobs (marked by arrows) along the plasma sheets PS1 and PS2.} 
\label{fig2}
\end{figure*}

\begin{figure}
\centering{
\includegraphics[width=9cm]{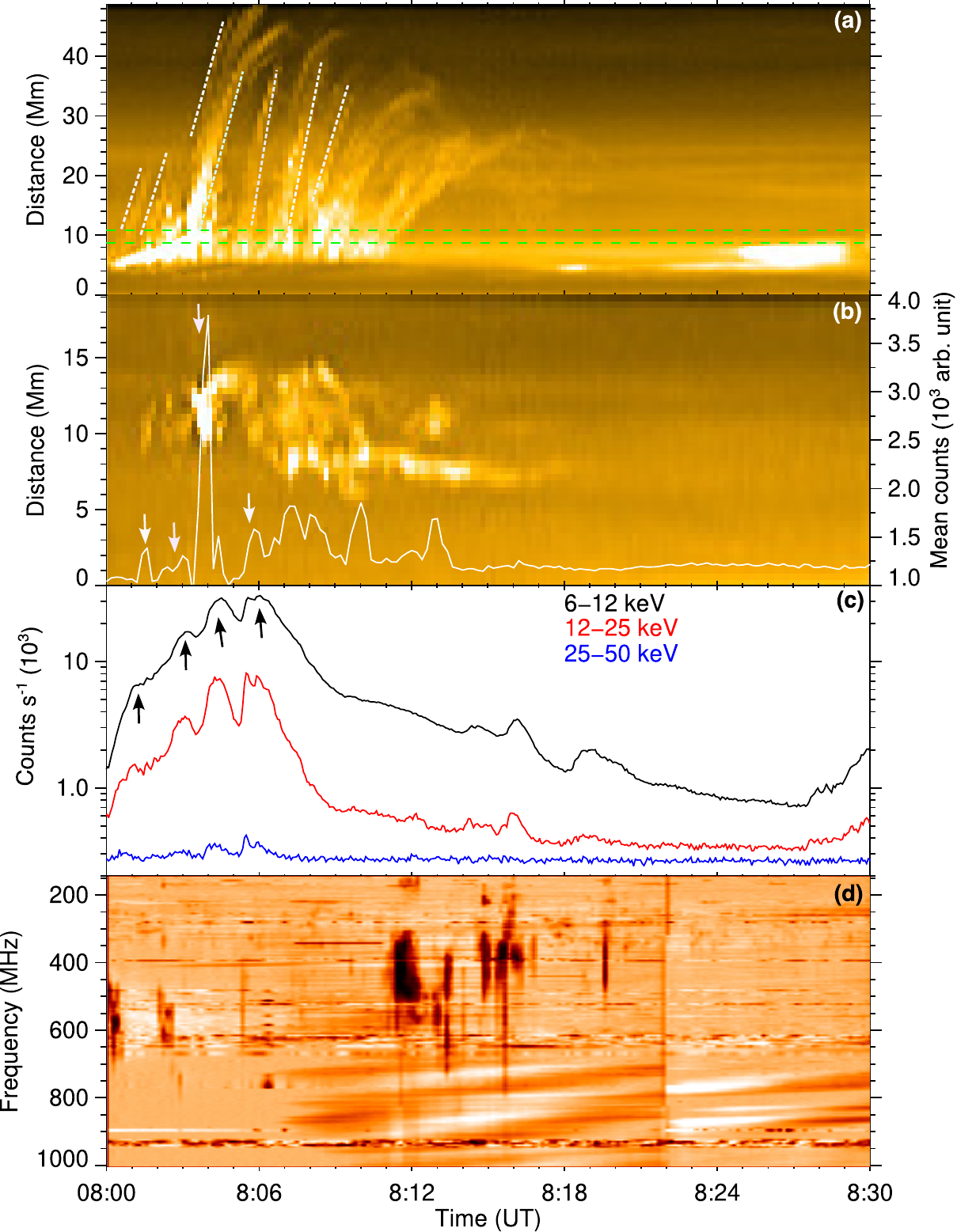}

}
\caption{{\bf X-ray and radio emissions associated with the ejection and merging of blobs during the first flare (C3.8) from 08:00-08:30 UT.} (a, b) Time-distance intensity plot along slices P1Q1 and P2Q2 using AIA 171~{\AA} images. The dashed lines represent the tracks used to estimate the speeds of the upward-moving blobs. The white curve represents the mean intensity (arbitrary units) extracted between the two green dashed lines in panel (a). (c) Fermi GBM counts in the 6-12, 12-25, and 25-50 keV energy bands. (d) ORFEES radio dynamic spectrum (144-1004 MHz).} 
\label{fig3}
\end{figure}
\begin{figure*}
\centering{
\includegraphics[width=18cm]{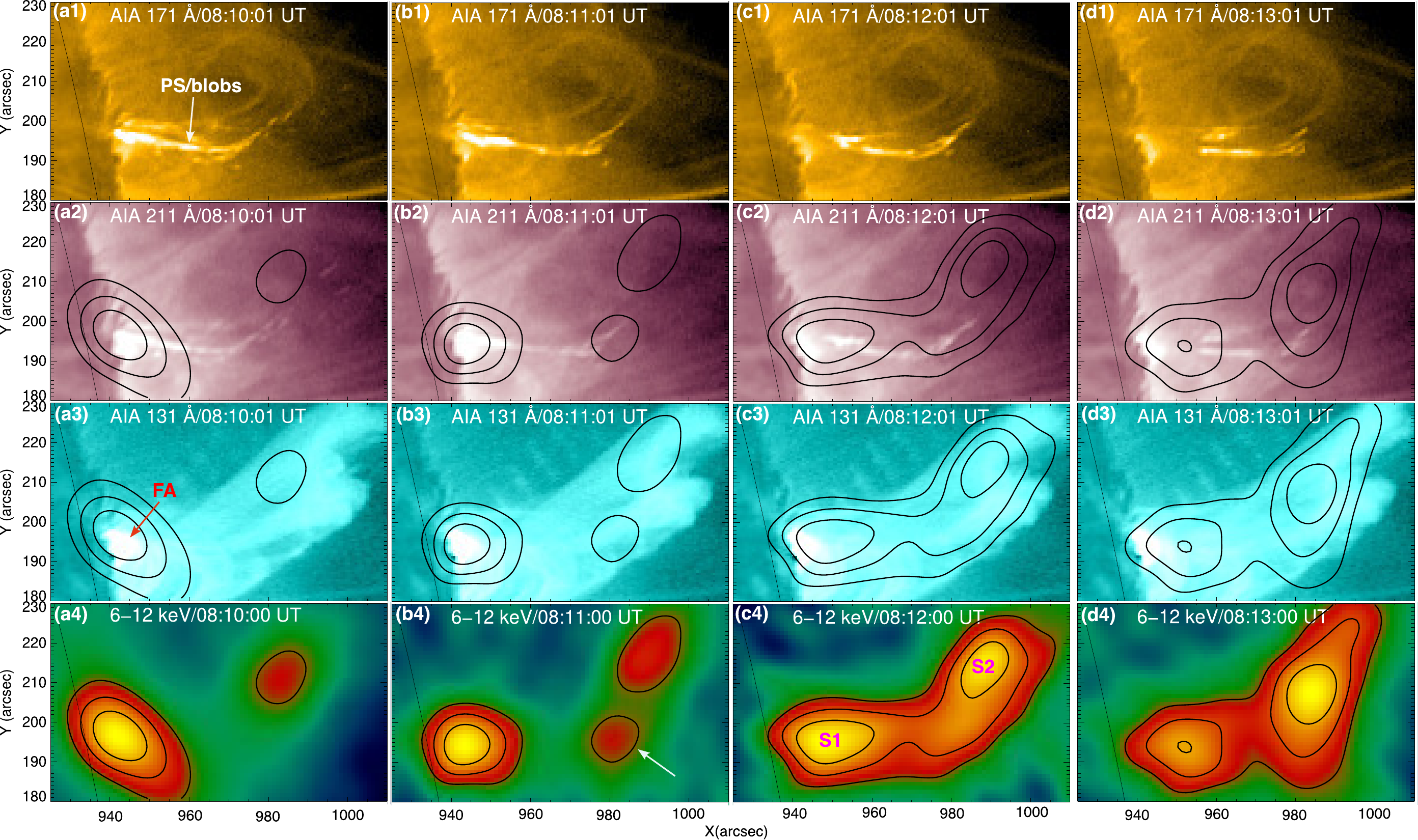}
}
\caption{{\bf Evolution of X-ray sources (6-12 keV) during the first flare (C3.8) from 08:10-08:13 UT.} AIA 171, 211, and 131~{\AA} images during the QPPs detected in the decay phase of the flare. RHESSI images in the 6-12 keV energy band overlaid on AIA 211 and 131~{\AA} images. The contour levels are 40$\%$, 60$\%$, 80$\%$ of the peak intensity. S1 and S2 represent double coronal sources.  PS=plasma sheet, FA=flare arcade.} 
\label{fig4}
\end{figure*}

\begin{figure*}
\centering{
\includegraphics[width=18cm]{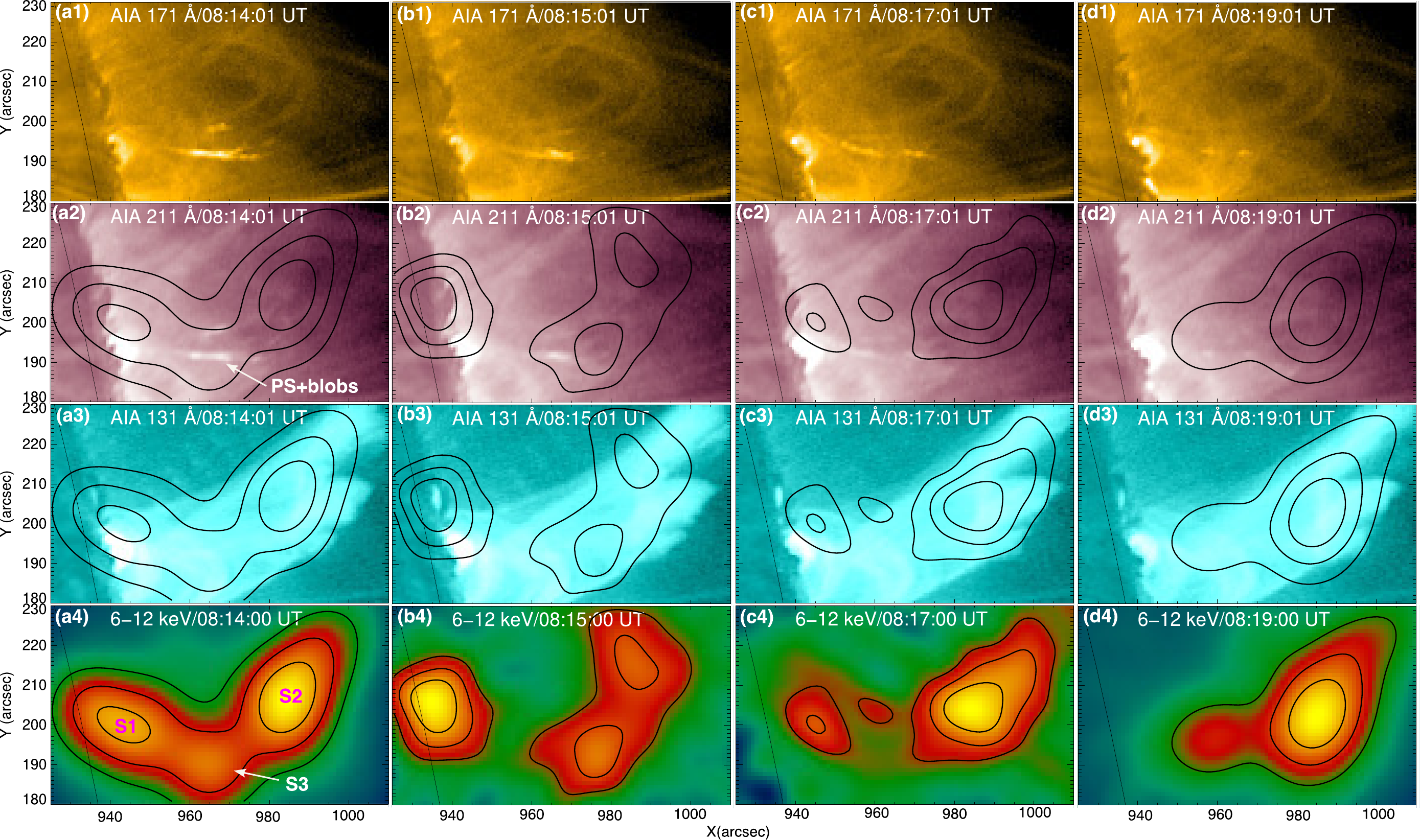}
}
\caption{{\bf Evolution of X-ray sources (6-12 keV) during the first flare (C3.8) from 08:14-08:19 UT.} AIA 171, 211, and 131~{\AA} images during the QPPs detected in the decay phase of the flare. RHESSI images in the 6-12 keV energy band overlaid on AIA 211 and 131~{\AA} images. The contour levels are 40$\%$, 60$\%$, 80$\%$ of the peak intensity. S1 and S2 represent double coronal sources. S3 is the faint source that appeared between S1 and S2.} 
\label{fig5}
\end{figure*}
\begin{figure*}
\centering{
\includegraphics[width=17cm]{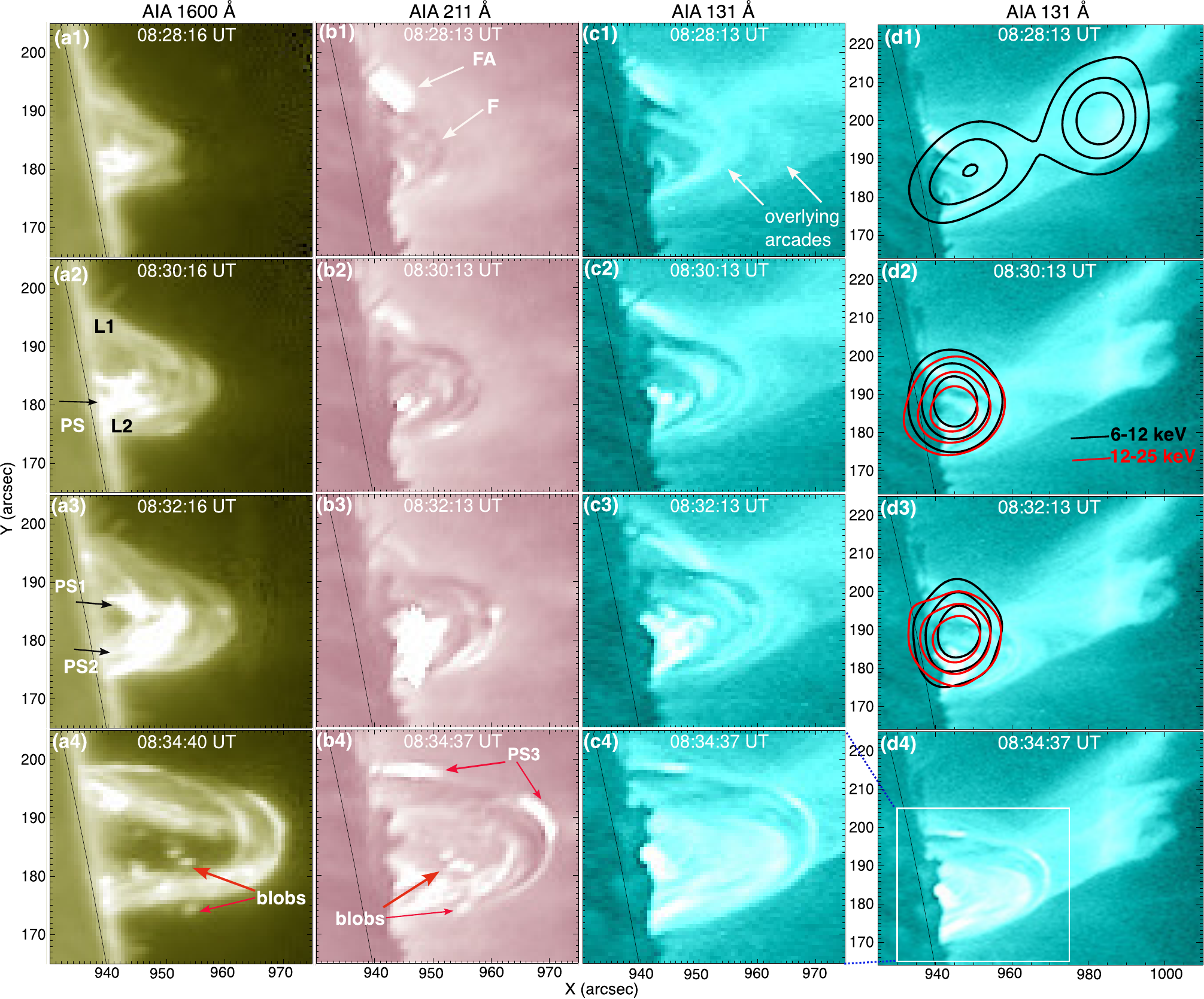}
}
\caption{{\bf Initiation of the second flare (M1.1) associated with a filament (F) eruption (08:28-08:34 UT).} (a1-c4) AIA 1600, 211, and 131~{\AA} images during the first stage of energy release. L1 and L2 represent the legs of the erupting filament. FA=flare arcade from the previous C3.8 flare, PS1 and PS2=flare plasma/current sheets along the legs of the flux rope, PS3=plasma/current sheet at the leading edge of the flux rope. (d1-d4) RHESSI images in 6-12 (black) and 12-25 (red) keV energy bands overlaid on AIA 131~{\AA} images. The contour levels are 40$\%$, 60$\%$, 80$\%$ of the peak intensity.} 
\label{fig6}
\end{figure*}
\begin{figure*}
\centering{
\includegraphics[width=16cm]{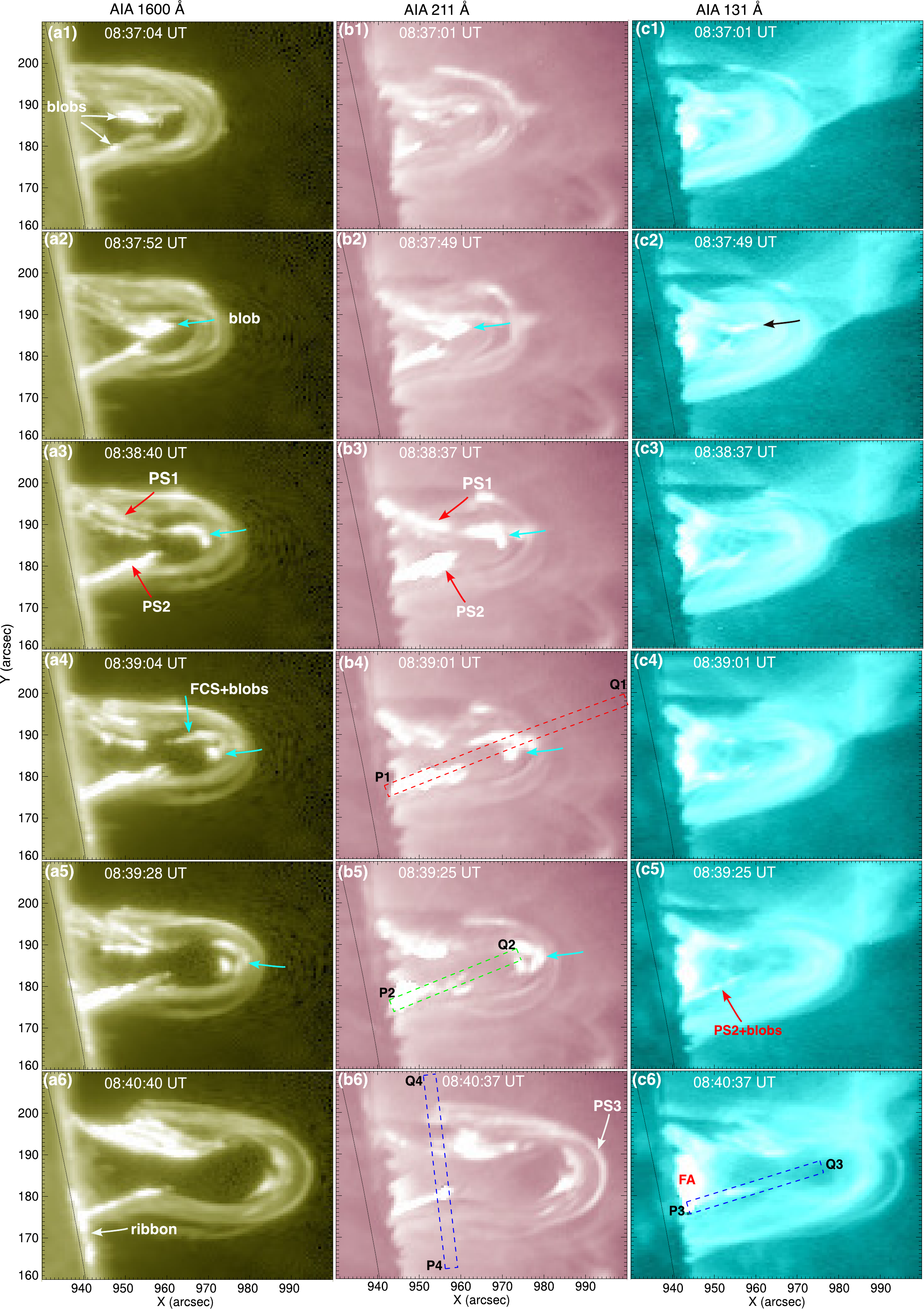}
}
\caption{{\bf Evolution of plasma sheets and blobs during the second flare (M1.1) from 08:37-08:40 UT.} (a1-c6) AIA 1600, 211, and 131~{\AA} images during the second stage of energy release. FA=flare arcade, PS1 $\&$ PS2 (red arrows)=flare plasma/current sheets along the legs of the flux rope, PS3=plasma/current sheet at the leading edge of the flux rope. Cyan arrows indicate the upward-moving blobs in the flare plasma/current sheet (FCS) below the rising flux rope. P1Q1, P2Q2, P3Q3, and P4Q4 are the slices used to create the time-distance intensity plots shown in Figure \ref{fig8}.} 
\label{fig7}
\end{figure*}
\begin{figure*}
\centering{
\includegraphics[width=16cm]{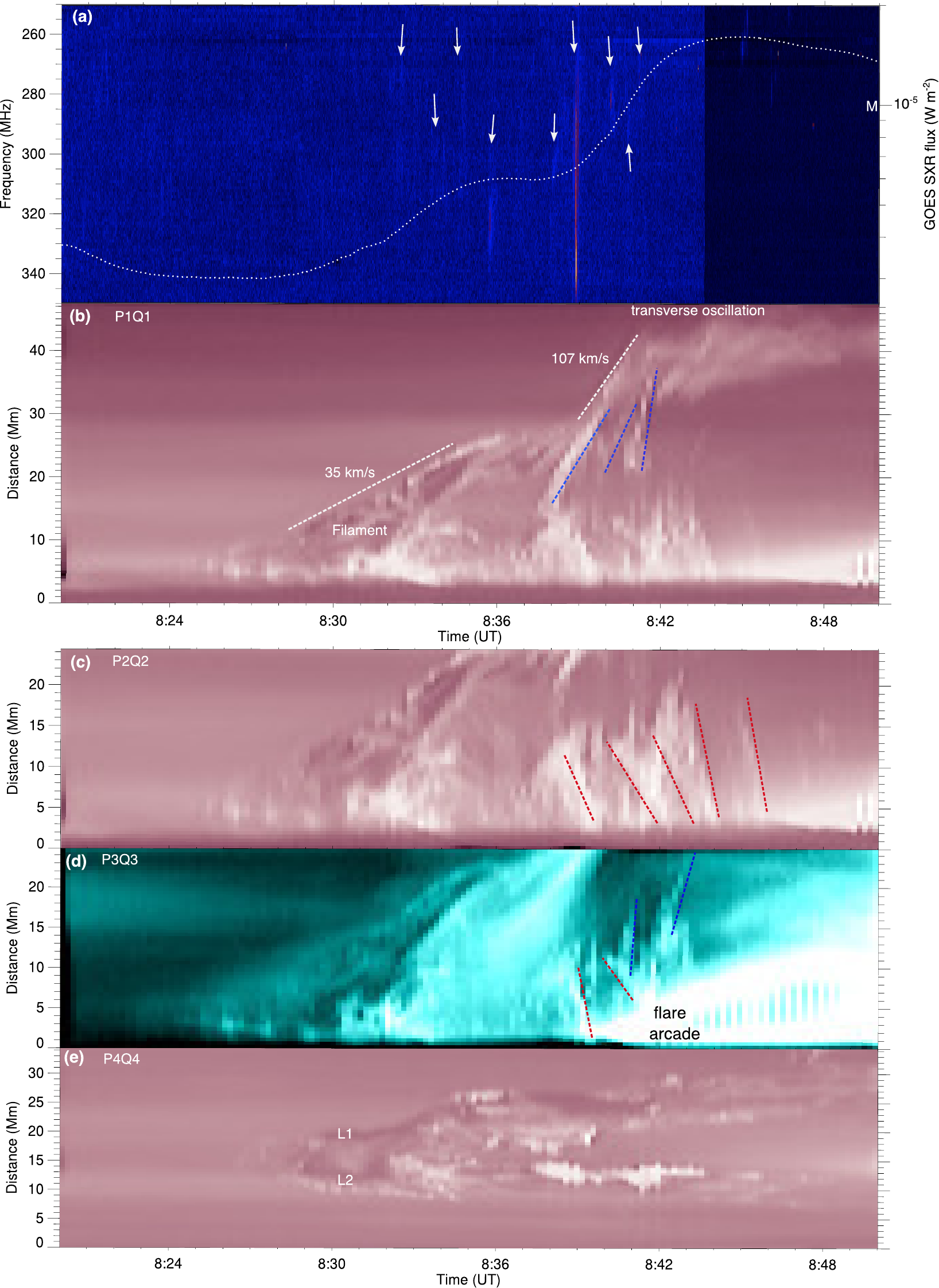}
}
\caption{{\bf Formation and ejection of upward/downward moving blobs and associated decimetric/metric radio bursts during the second flare (M1.1).} (a) e-Callisto dynamic radio spectrum (250-350 MHz) from the KRIM station (Crimean Astrophysical Observatory). The dotted curve is the GOES soft X-ray flux profile in the 1-8~{\AA} channel. (b-e) TD intensity plots along slices P1Q1, P2Q2, P3Q3, and P4Q4 (marked in Figure \ref{fig7}) using AIA 211 and 131~{\AA} images. The blue dashed lines indicate the upward-moving blobs (panels b \& d) while the red dashed lines (panels c, d) represent the downward-moving blobs. L1 and L2 are the legs of the erupting flux rope. The white arrows point to faint Type III bursts and one strong burst around 08:39 UT. } 
\label{fig8}
\end{figure*}
\begin{figure*}
\centering{
\includegraphics[width=18cm]{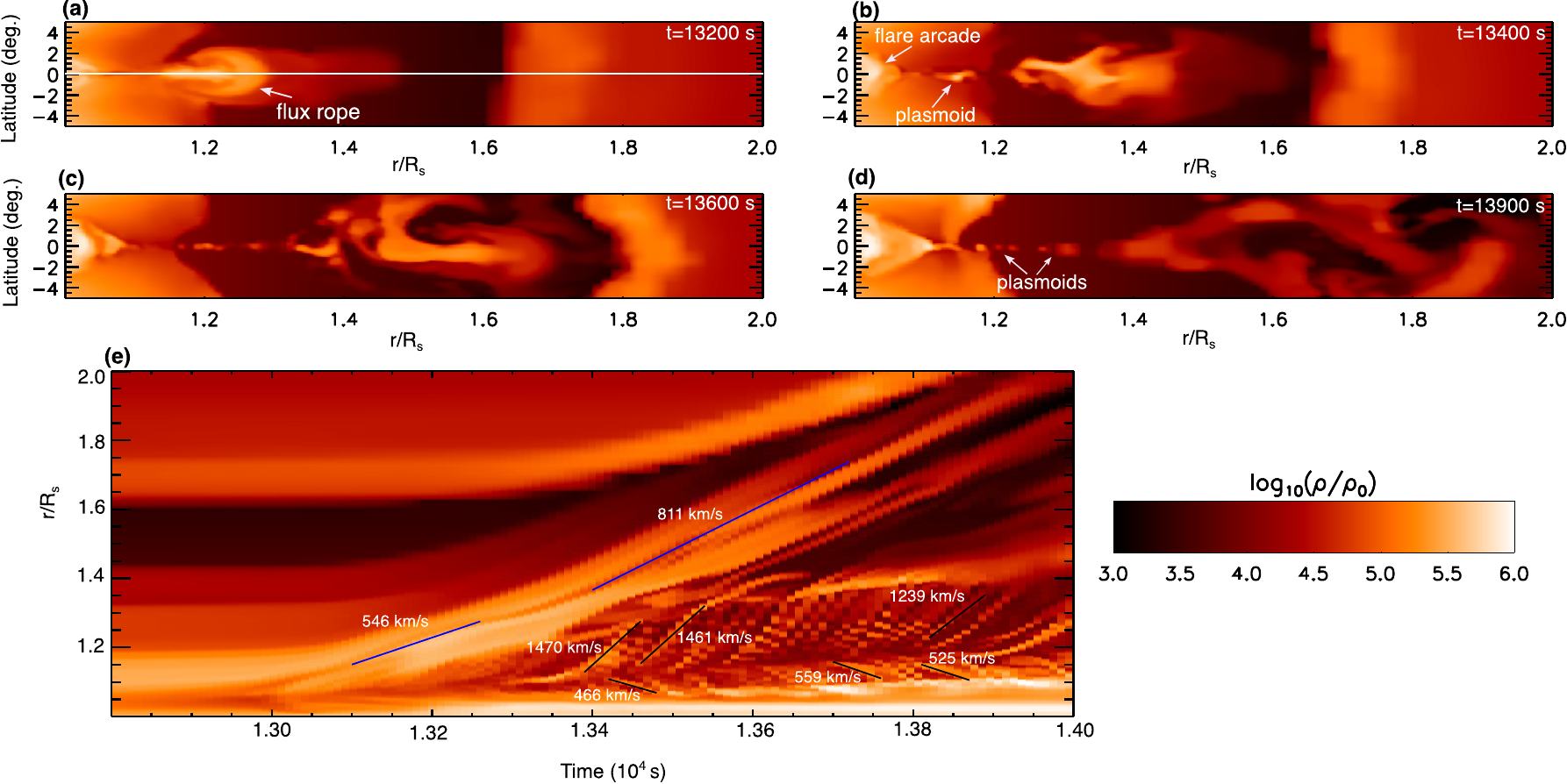}
}
\caption{{\bf 3D MHD simulation of formation and evolution of multiple plasmoids in a flare current sheet underneath an erupting flux rope.} (a-d) Density images at four selected intervals showing the temporal evolution of the erupting flux rope, flare arcade, and plasmoids. (e) TD density plot along a slice marked in panel (a) (from Dahlin et al., in preparation).} 
\label{fig9}
\end{figure*}

\section{OBSERVATIONS AND RESULTS}\label{obs}
We analyzed SDO/AIA \citep{lemen2012} full-disk images of the Sun (field-of-view $\approx$ 1.3~$R_\sun$) with a spatial resolution of 1.5$\arcsec$ (0.6$\arcsec$~pixel$^{-1}$) and a cadence of 12~s, in the following channels: 1600~\AA\ (\ion{C}{4}+continuum, temperature $T\approx 0.1$~MK, 5000~K), 304~\AA\ (\ion{He}{2}, temperature $T\approx 0.05$~MK), 171~\AA\ (\ion{Fe}{9}, $T\approx 0.7$~MK), 211~\AA\ (\ion{Fe}{14}, $T\approx2$~MK), 193~\AA\ (\ion{Fe}{12}, \ion{Fe}{24}, $T\approx  1.2$~MK and $\approx 20$~MK), and 131~\AA\ (\ion{Fe}{8}, \ion{Fe}{21}, \ion{Fe}{23}, i.e., $T\approx$ 0.4, 10, 16 MK) images. The 3D noise-gating technique \citep{deforest2017} was applied to denoise the SDO/AIA images.

To investigate the particle acceleration sites during the flares associated with these stalled flux-rope eruptions, we used X-ray lightcurves and images from the Reuven Ramaty High Energy Solar Spectroscopic Imager (RHESSI; \citet{Lin2002}). We used the CLEAN algorithm \citep{Aschwanden2004} to reconstruct the X-ray image, with an integration time of 60 s in the 6-12 keV and 12-25 keV energy channels using detectors 3 and 5-9. RHESSI missed the impulsive phases of the flares; therefore, we also analyzed Fermi Gamma-ray Burst Monitor (GBM; \citealt{meegan2009}) light curves during the flares in the 3-6, 6-12, and 12-25 keV ranges. We utilized dynamic radio spectra obtained by the Radio Solar Telescope Network (RSTN) Learmonth Radio Observatory, e-Callisto (extended-Compound Astronomical Low-frequency Low-cost Instrument for Spectroscopy and Transportable Observatory:\citep{callisto2009}), ORFEES (Observations Radiospectrographiques pour FEDOME et l'Etude des Eruptions Solaires: \citealt{hamini2021}) for metric/decimetric emission from the low corona, and WIND/WAVES \citep{bougeret1995} for the interplanetary medium.

 The active region NOAA 12322 was located on the west limb (N11W91) on 22 April 22 2015. The HMI magnetogram at 14:15 UT on 20 April, two days before the flares, shows the bipolar AR (inside the white oval) with a $\beta$ magnetic configuration (Figure \ref{fig1}(b)). The AIA 171 {\AA} image shows the overlying loops in the active region above the limb (Figure \ref{fig1}(a)). 
 We studied two limb flares (C3.8, M1.1) and the associated flux-rope eruption from AR 12322 on 22 April 2015. According to the GOES soft X-ray flux profile, the first flare (C3.8) started at 07:59 UT, peaked at 08:06 UT, and ended at about 08:24 UT. The second flare (M1.1) started at 08:28 UT, peaked at 08:44 UT, and ended around 08:58 UT (Figure \ref{fig1}(c)). The second flare shows two stages of energy release in the soft X-ray light curve. Both flares were nearly homologous and occurred successively from the same polarity inversion line (PIL) of the AR. The GOES soft X-ray flux derivative (1-8 {\AA}), which serves as a proxy for hard X-ray flux (i.e., Neupert effect; \citealt{neupert1968}), reveals quasi-periodic pulsations during both flares (Figure \ref{fig1}(d)). RHESSI partially observed these flares and missed the impulsive phases, covering the intervals 08:10-08:34 UT and 08:44-09:00 UT (Figure \ref{fig1}(e)). There is a strong correlation between the soft X-ray flux derivative and the observed hard X-ray bursts during both flares/eruptions, as expected. No CME\footnote{\url{https://cdaw.gsfc.nasa.gov/CME_list/index.html}} was detected in the LASCO C2 coronagraph images during these flares. The Learmonth (25-180 MHz) and Wind/WAVES dynamic radio spectra do not show any radio bursts (type II or III) associated with these events.
 
\subsection{Event $\#$1 (E1)}
AIA images during the first flare reveal the appearance of a flux rope, which was detected only in the hot channels (AIA 94 and 131 {~\AA}; (Figure \ref{fig2}(a1-a4)). One leg (L1) of the flux rope is close to the flare arcade, while the other leg extends behind the limb (Figure \ref{fig2}(a4)). A bright plasma sheet (PS) above the flare arcade appeared between these two legs (Figure \ref{fig2}(a3, a4)). The flux rope was confined and did not produce a successful eruption (see Movie S1). 

The blobs were best seen in the AIA 171, 193, and 211~{\AA} channels; therefore, we show a sequence of AIA 171~{\AA} images (zoomed view of the region marked in Figure \ref{fig2}(a1) and Movie S2) covering the bright plasma sheet and blobs. AIA 171~{\AA} images reveal the ejection of multiple plasma blobs (marked by arrows) along the bright plasma sheet above the flare arcade during 08:00-08:11 UT (Figure \ref{fig2}(b1-b4, c1-c4)).  During 08:04-08:10 UT the bright sheet has two components (PS1 and PS2) in which the plasma blobs propagate (Figure \ref{fig2}(b4-c3)). After 08:10 UT we observed only a single plasma sheet. The size of the blobs ranges from 2-3 arcseconds. 

To measure the temporal evolution and kinematics of the blobs formed in the flare plasma sheet, we created time-distance (TD) EUV intensity plots along slices P1Q1 and P2Q2 (marked in Figure \ref{fig2}(b1) ) using AIA 171~{\AA} images during the first flare (C3.8; 08:00-08:30 UT). The TD intensity plot along P1Q1 (along the plasma sheet) shows the ejection of multiple blobs above the flare arcade during the flare impulsive phase (08:00-08:11 UT;  Figure \ref{fig3}(a)). The speeds of the upward-moving blobs along the tracked paths are 228, 203, 208, 295, 323, 370, and 210 \kms. The TD intensity plot along P2Q2 (across the plasma sheet) reveals the blobs passing through the slit (Figure \ref{fig3}(b)). The blobs were detected in the plasma sheet until 08:20 UT and disappeared afterward. The white curve on the right y-axis represents the average intensity of the blobs, which is extracted between the two horizontal dashed lines in panel (a).

The Fermi GBM light curve shows the X-ray emission in the 6-12, 12-25, and 25-50 keV channels (Figure \ref{fig3}(c)). The X-ray emission reveals QPPs in the 6-12 and 12-25 keV bands during 08:00-08:20 UT, with only weak emission in the 25-50 keV band (08:00-08:06 UT). The ejection of blobs is nearly associated with the X-ray emission peaks (marked by arrows) in the 6-12 and 12-25 keV bands during the flare impulsive phase (four cycles during 08:00-08:08 UT, see Movie S2). Interestingly, QPPs also appear in the same X-ray bands during the flare decay phase, 08:11-08:20 UT. The dynamic radio spectrum shows nearly simultaneous decimetric/metric radio bursts (800-150 MHz) with the X-ray peaks during the flare impulsive and decay phases (Figure \ref{fig3}(d)). The radio/X-ray emission stopped at 08:20, coinciding with the disappearance of the blobs in the plasma sheet.

We utilized e-Callisto dynamic radio spectra (250-350 MHz) from the KRIM station (Crimean Astrophysical Observatory) to investigate the metric/decimetric bursts during the flares. The fine structure of radio bursts (pulsating structures, marked by arrows) is best seen in the e-Callisto dynamic spectra (Figure \ref{app-fig1}). The radio pulsating structures, detected in metric/decimetric frequencies, reveal strong correlation with the X-ray emission in the 12-25 keV band during the first flare (C3.8). Some of the radio bursts also show multiple substructures at 08:02, 08:12, and 08:15 UT (Figure \ref{app-fig1}(a,b)).

To observe the evolution of the RHESSI X-ray sources during the QPPs, we reconstructed RHESSI images in the 6-12 keV range using the CLEAN algorithm. The AIA 171 and 211 \AA\ images exhibit a bright plasma sheet and blobs during 08:10-08:13 UT (Figure \ref{fig4}), while the AIA 131~{\AA} images show the bright flare arcade (FA) and flux rope. The RHESSI image at 08:10 UT reveals two sources above the limb (Figure \ref{fig4}(a4)). The lower source coincides with the flare arcade, while the upper source aligns with the top portion of the flux rope. Notably, we detected simultaneous X-ray (6-12 keV) and radio bursts during 08:11:30-08:12:30 UT (see Figure \ref{app-fig1}(a)). At 08:11 UT, the RHESSI image shows an additional source (marked by an arrow, Figure \ref{fig4}(b4)), which is spatially aligned with the top part of the plasma sheet with blobs. Later, at 08:12 UT, we observed an intensity increase and extension of both the upper and lower sources (S1, S2).

At 08:14 UT, we observed another faint source, S3, between S1 and S2 (Figure \ref{fig5}(a4)). One minute later, the source had moved up about 12 arcsecs in the RHESSI image. The estimated speed of S3 is around 150 \kms. Simultaneously, we noticed a decrease in the height of S1 (Figure \ref{fig5}(b4)). RHESSI contours overlaid on AIA images reveal that source S3 coincides with blobs near the apex of the plasma sheet. Source S1 fades as its height increases during 08:17-08:19 UT, while source S2 remains bright (Figure \ref{fig5}(c4,d4)). The source S1 disappeared after 08:20 UT and source S2 also faded gradually (see Movie S3) during the decay phase of the flare.

\subsection{Event $\#$2 (E2)}
The second flare (M1.1) began at 08:28 UT, peaked at 08:44 UT, and ended around 08:58 UT. The GOES flux shows two stages of energy release during this flare. During the first stage of energy release (08:28-08:36 UT), AIA images (1600, 211, and 131~{\AA}) reveal the activation of a filament close to the flare arcade from the previous C3.8 flare (Figure \ref{fig6}). The slow rise of the filament begins with brightening between its legs. The AIA images and associated animation (Movie S4) depict the kinking of the slowly rising filament (F) along with the formation of a bright plasma/current sheet (PS) between its legs (Figure \ref{fig6}(a2,b2,c2)). As for the first event, two bright plasma sheets (PS1 and PS2 aligned along both legs) are observed around 08:32 UT (Figure \ref{fig6}(a3)).  A bright plasma sheet (PS3) appears at the leading edge of the erupting filament (Figure \ref{fig6}(a4,b4,c4)) around 08:34 UT, during the encounter of the filament with the overlying arcades marked in Figure \ref{fig6}(c1). Multiple plasma blobs (width$\approx$2-3 arcsecs) were detected in the flare plasma sheet beneath the erupting filament at about 8:34 UT (Figure \ref{fig6}(a4,b4)), leading to fragmentation of the sheet. Simultaneous AIA 211 and 131~{\AA} images also reveal a bright plasma sheet between the legs of the erupting filament.

RHESSI only observed the first stage of energy release until $\approx08:34$ UT. Unfortunately, it missed the impulsive phase of the second stage of energy release (M1.1 flare). The RHESSI X-ray source contours overlaid on the AIA 131~{\AA} image at 08:28:13 UT indicate the appearance of a lower source associated with the activation of the filament. Note that the higher source is associated with the hot emission from the halted flux rope (Figure \ref{fig6}(d1)), observed during the previous C3.8 flare. Later, we only see a single X-ray source (6-12 and 12-25 keV) located between the legs of the erupting filament (Figure \ref{fig6}(d2, d3)) during 08:30-08:33 UT (Movie S5).

The filament appears to be stable for a while and is confined during the decay phase of the first energy release (08:34-08:37 UT). During the impulsive phase of the second energy release (08:38-08:40 UT), we observed two plasma sheets along the flux-rope legs (PS1, PS2) with blobs moving upward and downward within both sheets aligned along the legs of the filament (Figure \ref{fig7}). We also observed upward-moving blobs in the flare current sheet (FCS), which merge, collide with the lower part of the flux rope, and change their orientation towards the southern leg (cyan arrow in Figures \ref{fig7}(a2-a5) and \ref{fig7}(b2-b5)). AIA 131~{\AA} images reveal bright plasma sheets along the legs, the formation and ejection of multiple blobs, and the flare arcade (FA) during the impulsive phase of the flare (Figure \ref{fig7}(c1-c6) and Movie S6). AIA 1600~{\AA} images reveal a ribbon near the southern footpoint of the flux rope (Figure \ref{fig7}); the second ribbon is likely occulted behind the limb. The flux rope stopped nearly at the same height (about 50$\arcsec$ above the limb) as the flux rope during the first flare (C3.8).

We created time-distance intensity plots from AIA 211 and 131~{\AA} images along slices P1Q1, P2Q2, P3Q3, and P4Q4 to determine the kinematics of the erupting filament and associated upward/downward moving blobs in the flare current sheet during 08:20-08:50 UT. The dynamic radio spectrum from e-Callisto reveals faint decimetric/metric radio bursts (250-350 MHz) occurring prior to and during the impulsive phase of the M1.1 flare (Figure \ref{fig8}(a)). The AIA 211~{\AA} time-distance intensity plot along slice P1Q1 captures the erupting filament and plasma sheet (Figure \ref{fig8}(b)). The filament rises slowly at about 35 \kms during 08:25-08:35 UT. Bright plasma sheets were observed along the legs of the erupting flux rope (PS1/PS2) and at the leading edge (PS3). The plasma sheet at the leading edge disappeared between 08:36 and 08:38 UT. A rapid rise of the flux rope at 107 \kms was observed during 08:39-08:41 UT, during the fragmentation of the plasma sheet, accompanied by bidirectional blobs. A strong radio burst was detected at 08:39 UT during the merging of an upward-moving blob with the lower boundary of the flux rope. Multiple blobs were observed until 08:46 UT. The flux rope was confined at 08:41 UT and exhibited transverse (kink) oscillations with a period of about 2-3 minutes and an amplitude of approximately 5-6 Mm (Figure \ref{fig8}(a)). At least three cycles of transverse oscillation are clearly observed between 08:40 and 08:47 UT. Notably, the kinematics of the flux rope correlate with the GOES soft X-ray flux, indicating two stages of energy release. During both stages, the formation and ejection of multiple plasmoids within the double structure of the plasma sheets were observed, along with associated decimetric/metric radio bursts (see Movie S6). The speeds of the blobs were determined by tracking the most visible paths in the time-distance intensity plots. The speeds of multiple upward-moving blobs (blue dashed lines) were 134, 150, and 330 \kms along P1Q1 (Figure \ref{fig8}(b)) and 254 and 175 \kms along P3Q3 (Figure \ref{fig8}(d)). The speeds of downward-moving blobs were 150, 82, 123, 235, and 204 \kms along P2Q2 (Figure \ref{fig8}(c)) and 162 and 88 \kms along P3Q3 (Figure \ref{fig8}(d)). The time-distance intensity plot along P4Q4 shows the separation of the flux rope legs during the eruption (Figure \ref{fig8}(e)). The brightenings between the legs are due to the downward-moving blobs in the plasma sheets. The AIA 131~{\AA} channel reveals the appearance of a flare arcade at 08:40 onwards due to ongoing reconnection in the flare current sheet (Figure \ref{fig8}(d) and accompanying Movie S6).

\section{DISCUSSION}\label{discussion}

\subsection{Plasma/current sheets with multiple plasmoids}
We analyzed two flare events that occurred successively at the main polarity inversion line (PIL) within the same active region. The blobs observed in the plasma sheets during both flares are interpreted as plasmoids. For the first time, we observed a double structure of the plasma/current sheet with multiple propagating blobs below the erupting flux ropes. We interpret these blobs as plasmoids formed by reconnection in the flare current sheet. During the first flare (C3.8), we observed upward-moving reconnection outflows traced by multiple plasmoids (speed: 200-370 \kms) and the formation of a hot flux rope during the eruption. Prior to the second flare (M1.1), a filament within the flux rope rose slowly (35 \kms) associated with brightening between its legs. The flux rope showed evidence of kink instability and the formation of two bright plasma sheets (PS1 and PS2) on the inner surfaces of its legs, joined by a central flare current sheet. The speeds of the upward and downward moving plasmoids were 134-330 \kms and 82-235 \kms, respectively, with plasmoid sizes ranging from 2-3 arcseconds. The sizes and speeds of the plasmoids are consistent with previous observations \citep{takasao2012,kumar2013,kumar2019b,kumar2023}.

The observations of plasma/current sheets detected below an erupting flux rope are consistent with the predictions of a MHD simulation of a kink-unstable flux rope. Figure 4 in \citet{kliem2010} depicts the double structure of the plasma sheet (helical current sheet in cyan color) along the legs of the flux rope and a flare current sheet (red color) below the flux rope. This MHD simulation exhibited current sheets and associated reconnection in a failed/confined eruption, strikingly similar to the observed events described in \S3.  However, the spatial resolution of the \citet{kliem2010} simulation does not seem to be sufficient to resolve plasmoids in the helical and flare current sheets. In contrast, we detected a separation between the legs of the flux rope during the eruption but did not observe leg-leg reconnection as seen in the MHD simulation. As described in \citet{kliem2010}, the helical current sheet forms as a result of the kink instability of a flux rope, wrapping around the twisted legs of the rope. This structure is crucial for understanding the magnetic reconnection processes and energy release during stalled solar eruptions.

We interpret the plasmoid dynamics with a comparison to the results of an MHD simulation (Figure \ref{fig9}). This simulation, which was performed with the ARMS code \citep{devore2008}, modeled an eruptive flare with a high-resolution adaptive grid using the same configuration employed in \citet{dahlin2022} at a lower resolution (the lower resolution was employed in order to capture the entire flare evolution, which was not computationally feasible at the higher resolution). Many plasmoids were formed, the dynamics of which will be discussed in detail in Dahlin et al. (in preparation). Here we show a selection of longitudinal slice of the density (Figure \ref{fig9}(a-d)), showing plasma blobs that correspond to plasmoids below an erupting flux rope. 
The running-difference animation of synthetic white-light images (constructed from the ARMS density data following the method of \citealt{lynch2016}) provides a better view of bidirectional plasmoids (Movie S7b). A TD plot of the density in Figure~\ref{fig9}(e) illustrates the formation and ejection of multiple plasmoids (Movie S7a). The acceleration of the flux rope during the period of plasmoid proliferation, consistent with the observations, is clearly evident.
In the simulations, the plasmoid speeds in the flare current sheets are higher than the blob speeds in the observations. This is probably due to the difference in the localized Alfve\'n speed between the simulation and the observed activity.

\subsection{Double X-ray sources}
During the first flare (C3.8), we detected double coronal X-ray sources (6-12, 12-25 keV) located at both ends of the plasma sheet (i.e., below and above the reconnection site). The lower source was located above the flare arcade, while the higher source was observed near the reconnection outflow, where multiple plasmoids merged with the underside of the halted flux rope. The flux rope appearing in the hot channels was observed during magnetic reconnection at the flare plasma/current sheet. Previous observations from RHESSI \citep{sui2003,sui2004,liu2008} have interpreted the double coronal source as evidence of magnetic reconnection in the flare current sheet.  However, direct imaging of the plasma/current sheet with multiple plasmoids moving bidirectionally has not been reported previously to be simultaneous with double coronal X-ray sources. Here, we present a consistent picture of plasmoid-mediated reconnection in the flare current sheet associated with the formation of double coronal X-ray sources, thereby adding observational support to previous findings by RHESSI. In addition, we observed a faint X-ray source (6-12 keV) that appeared between the double X-ray sources (S1, S2) and was co-spatial with EUV plasmoids in the plasma sheet. The speeds of the upward-moving faint source and the EUV plasmoids are consistent.

During the initiation of the second flare, we observed a RHESSI X-ray source (6-12, 12-25 keV) located near the plasma sheet between the legs of the flux rope (Figure \ref{fig6}(d2,d3)). This is consistent with plasma heating and particle acceleration associated with the formation and ejection of plasmoids in a plasma sheet beneath an erupting flux rope. Previous observations have also revealed the formation of a hard X-ray source between the legs of a kink-unstable flux rope, interpreted as magnetic reconnection in a current sheet \citep{alexander2006,cho2009}. However, these observations did not show evidence for a plasma/current sheet or plasmoids. Our observations provide direct imaging of plasma/current sheets with multiple plasmoids, thereby supporting previous interpretations about the formation of X-ray sources between the legs of a kink-unstable flux rope.

RHESSI missed the impulsive phase of the second flare, so we do not have X-ray imaging for that interval. We observed decimetric radio bursts (evidence of electron injections) associated with the formation/ejection and coalescence of plasmoids into the trailing edge of the flux rope during the impulsive phase of the M1.1 flare.

The Fermi GBM X-ray spectra (6-30 keV energy, assuming that the emission is generated by a thin-target bremsstrahlung radiation process) at different intervals reveal the evolution of temperature, emission measure, and spectral index during plasmoid-mediated reconnection in both flares (Figures \ref{app-fig5} and \ref{app-fig6}). The estimated temperature and emission measure near the quasi-periodic peaks (five intervals mentioned in the Fermi spectra) during the first flare (C3.8) range from 10 to 15 MK, and 1.9 to 6.8$\times$10$^{47}$ cm$^{-3}$. The spectral index varies between 3.9 and 6.7. For the second flare (M1.1), the temperature and emission measure near two peaks were [12, 13.5] MK, and [30$\times$10$^{47}$, 23$\times$10$^{47}$ cm$^{-3}$]. The spectral index ($\delta$) varies from 7.1 to 4.3 (see Table 1).

Quasi-periodic pulsations in decimetric radio bursts associated with plasmoids in a flare current sheet are most likely due to plasma emission. This mechanism effectively explains the rapid, periodic variations and fine structures observed in the radio spectrum, as it directly involves the interaction of accelerated electrons with the ambient plasma in the highly dynamic environment of the flare current sheet.
The observed frequencies of 300-800 MHz correspond to plasma densities (using $f_p=8980 \sqrt n_e$, where f$_p$ is the plasma frequency in Hz and n$_e$ is the electron density in cm$^{-3}$) ranging from approximately 1.1$\times$10$^9$ to $0.8\times10^{10}$ cm$^{-3}$ (assuming fundamental emission). The comparison of radio spectra from ORFEES (Figure \ref{fig3}(d)) and Callisto/KRIM (Figure \ref{app-fig1}) suggests that the ORFEES pattern (some of the type-III burst-like features) is likely the second harmonic of much fainter fundamental emission observed in the Callisto dynamic spectrum. If the radio emission for the frequency range of 500–800 MHz corresponds to the second harmonic (2f$_p$), the estimated plasma density will be approximately 0.7$\times$10$^9$ to 0.2$\times$10$^{10}$ cm$^{-3}$. These densities suggest that the emission source region should be located in the plasma sheet (i.e., upward-moving blobs merging into the flux rope).

Particle-in-Cell (PIC) simulations suggest that accelerated electron beams trapped within the helical structure of plasmoids, particularly in the lower-density shell, are believed to produce the DPS observed in radio bursts \citep{karlicky2011}.
The observations indicate that the decimetric pulsation is likely associated with electron beams injected during the coalescence of upward-moving blobs with the halted flux rope. The density of the plasma sheet (and blobs) is generally on the order of 10$^{10}$ cm$^{-3}$ (based on DEM analysis, Figures \ref{app-fig7} and \ref{app-fig8}), while the flux rope density is about 5.8$\times$10$^9$ cm$^{-3}$. The decimetric pulsations are temporally correlated with the appearance of the RHESSI upper source (S2), suggesting that upward-moving electron beams at different energies excite both X-ray (source S2) and decimetric radio emissions below the erupting flux rope (e.g., upward moving radio source in \citealt{pick2005}). The major X-ray emission is generally associated with the lower source S1 (i.e., downward-moving electron beams during reconnection).  

\subsection{QPPs associated with plasmoids}
The wavelet analysis \citep{torrence1998} of the X-ray light curve (12-25 keV) reveals a $\approx$100-s periodicity during the first flare (C3.8) (Figure \ref{app-fig3}). The quasiperiodic X-ray bursts closely match the decimetric pulsations (400-800 MHz) observed in the ORFEES dynamic spectrum (Figure \ref{fig3}(d)). This 100-s periodicity is correlated with the creation of multiple plasmoids in the flare current sheet. 

Moreover, the high resolution e-Callisto dynamic spectra and Learmonth observatory (1-s cadence) radio flux profiles reveal a shorter period in decimetric bursts. The wavelet analysis of the radio flux densities reveal a $\approx$10-s periodicity during the first flare (C3.8) (Figure \ref{app-fig4}). The 10-s period may be attributed to an MHD wave process. As plasmoids interact and merge, they undergo MHD oscillations (e.g., sausage mode, \citealt{nakariakov2003}), leading to perturbations in the density and magnetic field within the plasmoid (e.g., see MHD simulation by \citealt{jelnek2017}). These oscillations can modulate the injection of electron beams into the surrounding plasma, leading to repetitive acceleration of electrons. The periodic injection of these electron beams, synchronized with the plasmoid oscillations, may result in the observed 10-s quasi-periodic pulses in the radio emission. The AIA temporal resolution (12 s) is not sufficient to detect these oscillations in the EUV images (10 s). 
The typical size of plasmoids in our observations is about 2-3 arcsecs. The speed of upward-moving plasmoids ranges from 134 to 330 \kms. Using the observed oscillation period of the decimetric radio bursts, we determine the ambient Alfve\'n speed using the formula P=L/V$_A$, where L is the size (width) of the plasmoids and V$_A$ is the Alfve\'n speed at the edge of a plasmoid. For an average plasmoid size of 2000 km and an oscillation period of 10 seconds, the estimated Alfve\'n speed is about 200 \kms. This value is consistent with the average speed of the plasmoids in the flare current sheet.

During the second flare (M1.1), the QPPs in radio decimetric bursts are characterized by a period of about 60 seconds (Figure \ref{fig8}(a)), which is roughly consistent with the frequency of ejected plasmoids in the plasma sheet (Figure \ref{fig8}(b,c)). The strong radio burst around 08:37 UT was correlated with the merging of the first upward-moving blob with the lower boundary of the flux rope. The other upward-moving plasmoids also merge with the flux rope and produce weaker decimetric radio bursts.

The AIA 171/211~{\AA} running-difference images do not reveal any evidence of an EUV wave (shock/fast-mode waves) during the QPPs. Therefore, we can rule out the modulation of quasi-periodic particle acceleration via propagating fast EUV waves. Previous observations have revealed QPPs associated with upward-propagating fast-mode waves during flare reconnection and associated radio bursts without plasmoids \citep{kumar2017}.

QPPs observed in both radio and X-ray emissions during solar flares are often linked to the dynamic processes occurring in the flare current sheet, particularly the formation and ejection of plasmoids. These plasmoids can periodically release energy as they move through the current sheet. The corresponding radio and X-ray QPPs arise from the acceleration of electrons and the associated emissions as plasmoids interact with the surrounding plasma and magnetic fields. The temporal correlation between radio and X-ray QPPs suggests a common driver, i.e., the repetitive nature of plasmoid formation/ejection and associated coalescence. 

\subsection{Successive or magnetically connected confined eruptions}
Both eruptions occurred successively along the same PIL, with no gap between the flares. The flare arcades appeared next to each other. During the first flare, the flux rope was observed only in the hot channels (94/131 \AA). However, plasmoids were seen in multiple AIA channels covering cool and warm/hot plasma. The second flare started with the eruption of a filament-carrying flux rope along with bright plasma sheets along the two legs. The flux ropes in both eruptions stopped at a similar height (about 45 Mm). The AIA 171~{\AA} image reveals systems of multiple overlying loops above the erupting flux ropes (Figure \ref{fig1}(a)). We speculate that the flux ropes were unable to overcome the overlying strapping field as required for escape. The presence of two bright plasma sheets lining the flux-rope legs suggests the existence of a helical current sheet in both stalled eruptions. During the second eruption, we also noticed the appearance of a plasma sheet (PS3) at the leading edge of the flux rope during its interaction with the overlying structures. PS3 disappeared rapidly during the kinking motion of the flux rope. Further acceleration of the flux rope was associated with plasmoid-mediated reconnection in the current sheets embedded in PS1 and PS2. The majority of the upward-moving plasmoids merged into the underside of the flux rope. The flux rope halted, and no current sheet was left behind it for further flare reconnection. The flux rope plasma drained back to the solar surface after the eruption stalled. Therefore, the kinking motion, along with the inability of the flux rope to escape from the strong overlying flux system, is likely responsible for the confinement of these successive eruptions. 

We have observed a similar confined eruption associated with a kink-unstable flux rope with a flare plasma sheet containing plasmoids  \citep{kumar2023}. However, our previous observations did not exhibit the double structure of the helical plasma sheet along the flux-rope legs. In this event, we also observed a kink oscillation (period$\approx$2-3 minutes, amplitude$\approx$5-6 Mm) of the stalled flux rope after its encounter with the overlying loops \citep{kumar2022}. 

The kinematics of the flux rope and the GOES soft X-ray flux are strongly correlated during the second flare (M1.1). The slow rise of the flux rope matches the slow rise of X-ray flux prior to the flare (Figure \ref{fig8}). The flux rope remains nearly stable for about 2-3 minutes during the dip in soft X-ray flux, which we interpret as a pause in flare reconnection. The subsequent acceleration of the flux rope is strongly correlated with flare reconnection associated with plasmoids, suggesting that flare reconnection plays a significant role in the acceleration of the flux rope in the low corona.

\section{CONCLUSION}\label{discussion}
We report the direct imaging of the formation and ejection of multiple plasmoids in flare current sheets beneath erupting flux ropes, along with QPPs in X-ray and radio wavelengths. During the M1.1 (second) flare, we observed kinking of the flux rope in both hot and cool AIA channels, along with the formation and ejection of bidirectional plasmoids in the flare plasma sheet. The current/plasma sheet is predominantly viewed edge-on in this event, allowing for a more direct comparison with the 3D MHD simulation. In contrast, during the C3.8 (first) flare, the plasma/current sheet appears more distorted and tilted, with the flux rope formation observed only in the hot AIA channels. Upward-moving plasmoids are clearly visible, while downward-moving plasmoids appear less prominent during the first flare.
These observations confirm (i) that the X-ray double coronal sources observed by RHESSI are located at both ends of the flare current sheet, and formed during plasmoid-mediated reconnection in the sheet; (ii) the faint transient source that appeared between the double coronal sources is most likely associated with upward-moving plasmoids;  (iii) the presence of a flare current sheet with double structure and multiple plasmoids, as predicted by an MHD simulation of kink-unstable flux rope \citep{kliem2010}; (iv) the formation of a plasma/current sheet at the leading edge of the kink-unstable flux rope during its encounter with the overlying flux system; (v) that the coalescence of upward-moving plasmoids at the underside of the flux rope is accompanied by decimetric radio bursts; and (vi) X-ray/radio QPPs (P=10 s, 100 s) are associated with ejection and coalescence of plasmoids in the flare plasma/current sheet. These findings enhance our understanding of plasma heating and the quasi-periodic acceleration of electrons via plasmoid-mediated reconnection in flare current sheets below erupting flux ropes. The production of energetic electrons through the ejection and coalescence of plasmoids in the plasma sheet has broad applicability to understanding fast magnetic reconnection in solar, heliospheric, and magnetospheric current sheets. The direct imaging of plasmoids and associated QPPs provides key insights into plasmoid-mediated magnetic reconnection and particle acceleration, supporting theoretical models of these processes during solar flares.  In the future, similar flare events along with simultaneous EUV, radio, and hard X-ray imaging will yield further insights into the electron acceleration sites associated with plasmoids during magnetic reconnection in solar eruptions.\\
\\

\noindent
{\bf {Acknowledgements}}\\
SDO is a mission for NASA's Living With a Star (LWS) program. 
This research was supported by NASA's Heliophysics Guest Investigator (\#80NSSC20K0265), supporting research (\#80NSSC24K0264), Internal Scientist Funding Model (H-ISFM) programs, and NSF SHINE grant (\#2229336).
 Wavelet software was provided by C. Torrence and G. Compo, and is available at \url{http://paos.colorado.edu/research/wavelets/}.\\
 \\


\bibliographystyle{aasjournal}
\bibliography{reference.bib}

\clearpage
\appendix
\counterwithin{figure}{section}
\section{Supplementary figures}
\begin{figure*}[htp]
\centering{
\includegraphics[width=13.5cm]{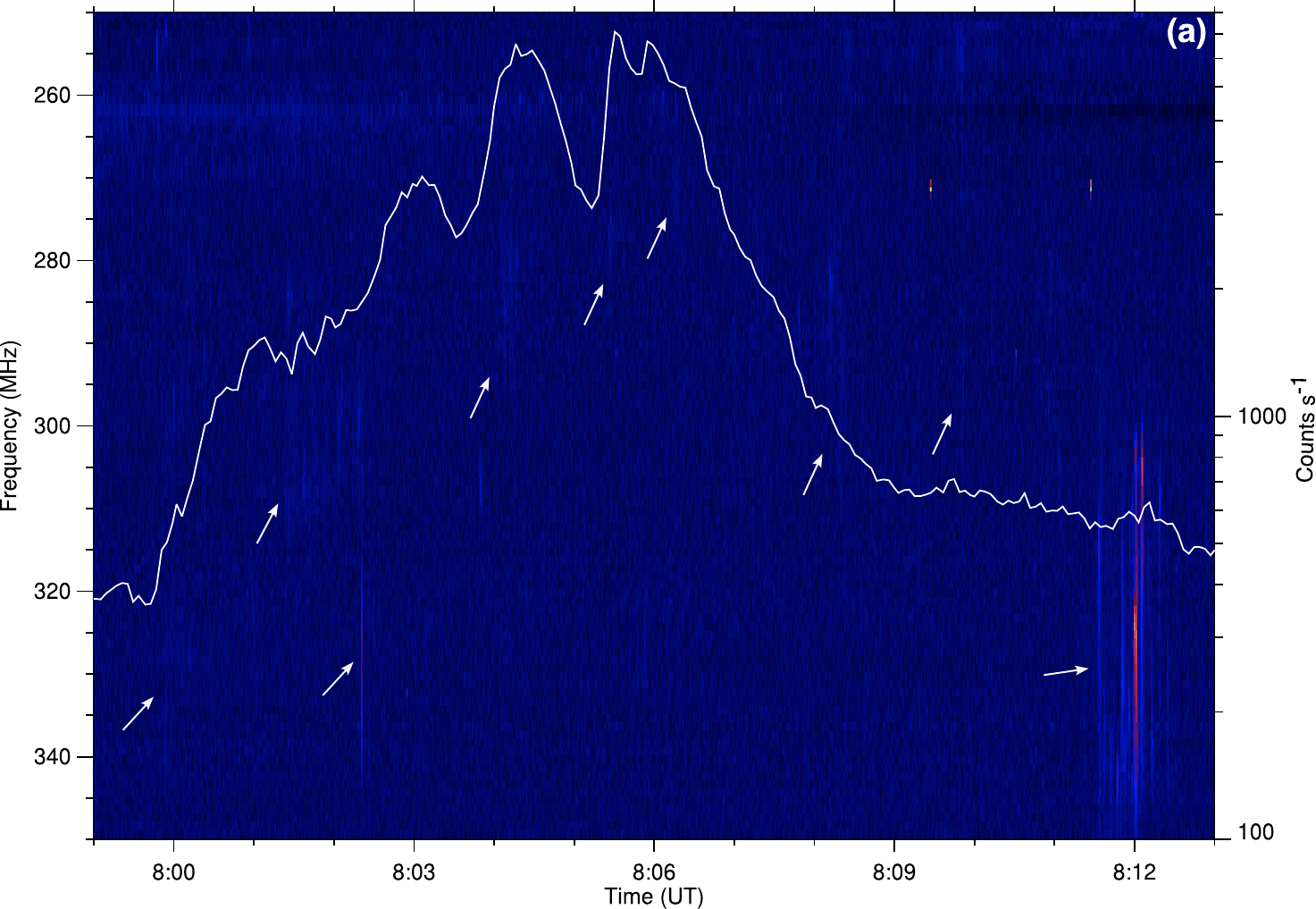}
\includegraphics[width=13.5cm]{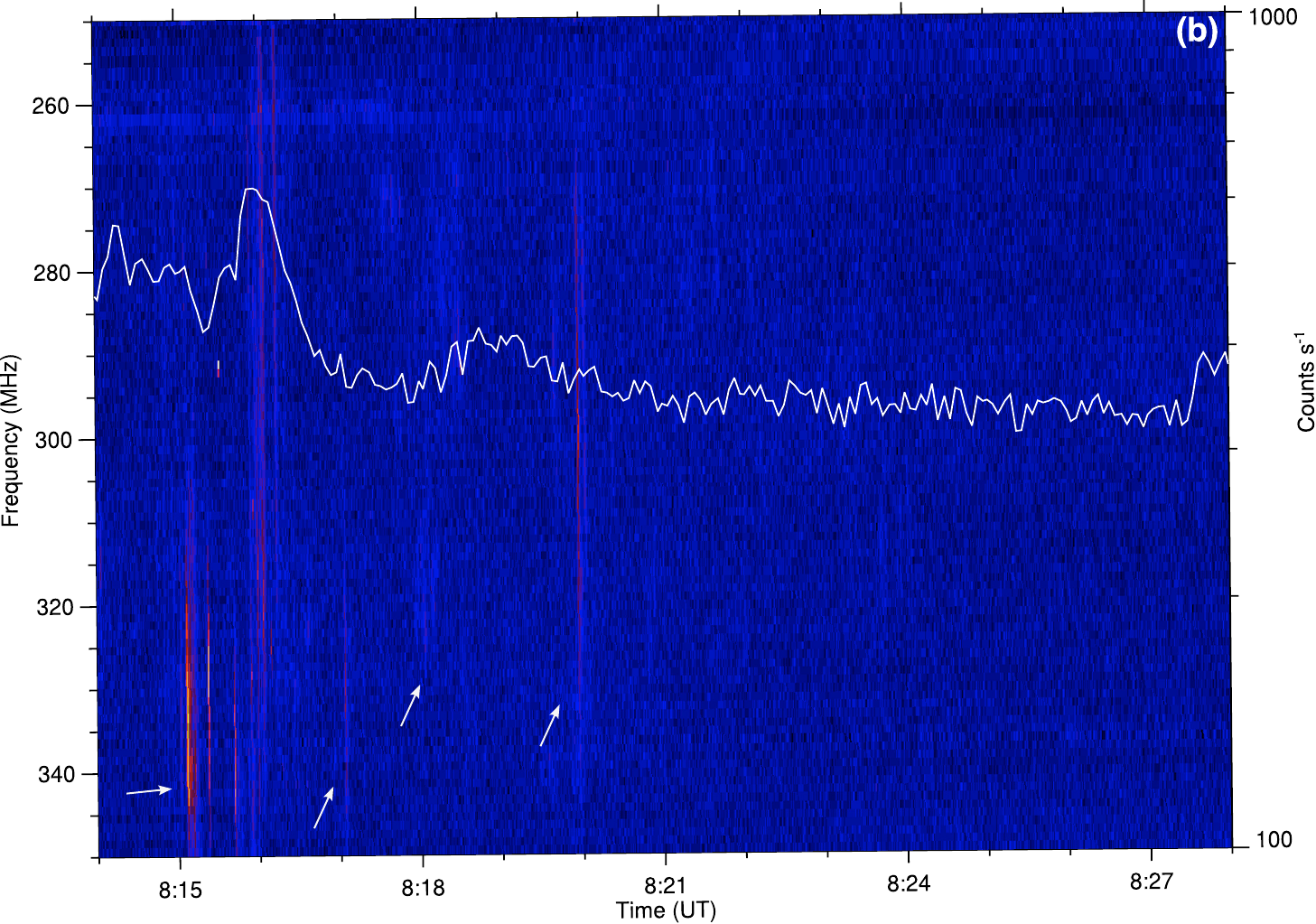}
}
\caption{{\bf Radio and X-ray emissions associated with the ejection and merging of blobs during the first flare (C3.8).} (a,b) e-Callisto dynamic radio spectrum (250-350 MHz) from the KRIM station (Crimean Astrophysical Observatory). The right Y-axis shows the Fermi GBM light curve (white) in the 12-25 keV energy band. The arrows indicate the radio bursts associated with the X-ray emissions.} 
\label{app-fig1}
\end{figure*}

\begin{figure*}
\centering{
\includegraphics[width=15cm]{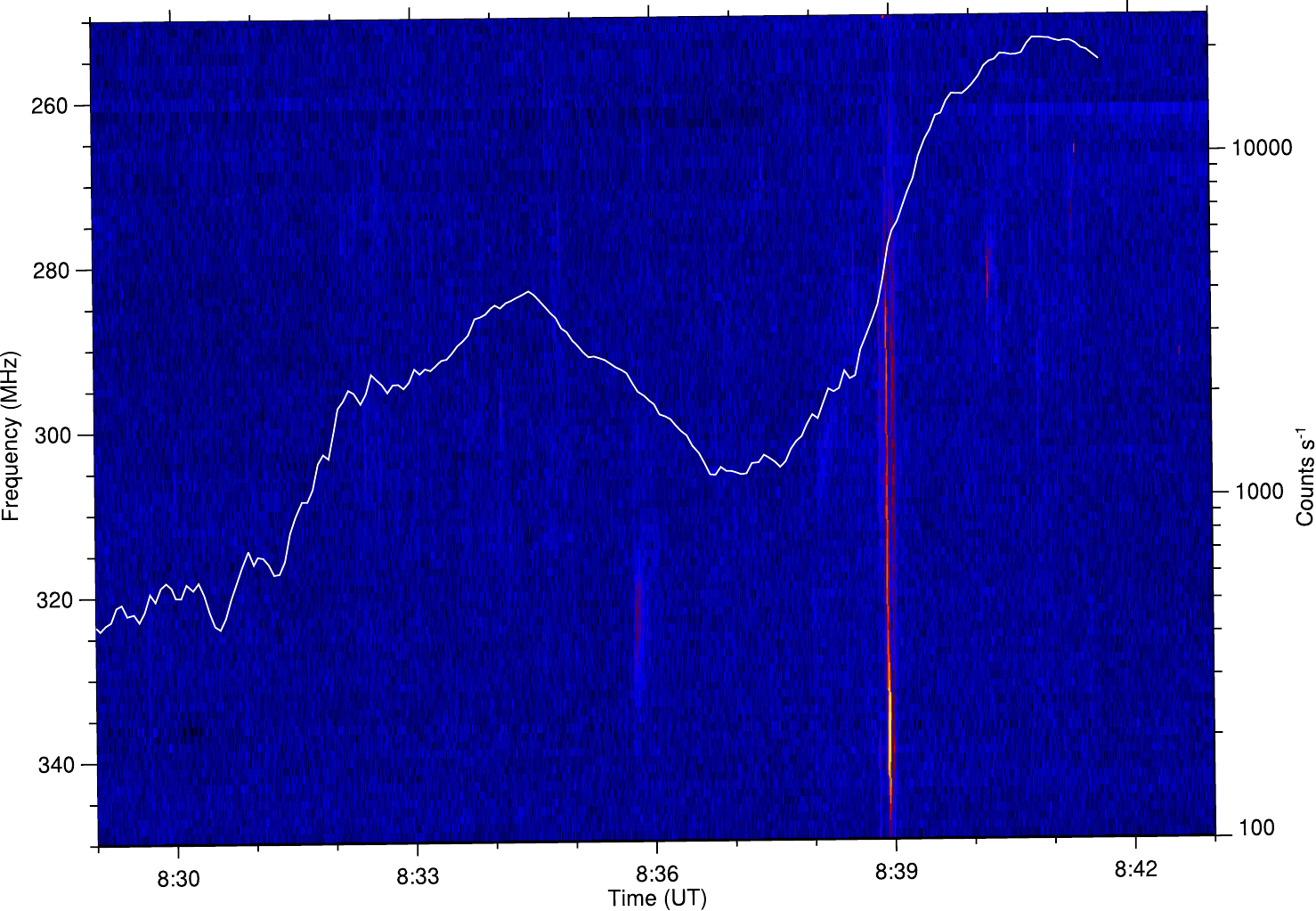}
}
\caption{{\bf Radio and X-ray emissions associated with the ejection and merging of blobs during the second flare (M1.1).} (a,b) e-Callisto dynamic radio spectrum (250-350 MHz) from the KRIM station (Crimean Astrophysical Observatory). The right Y-axis shows the Fermi GBM light curve (white) in the 12-25 keV energy band. Fermi GBM did not observe the decay phase of the flare (after 08:41:40 UT).} 
\label{app-fig2}
\end{figure*}

\begin{figure*}
\centering{
\includegraphics[width=18cm]{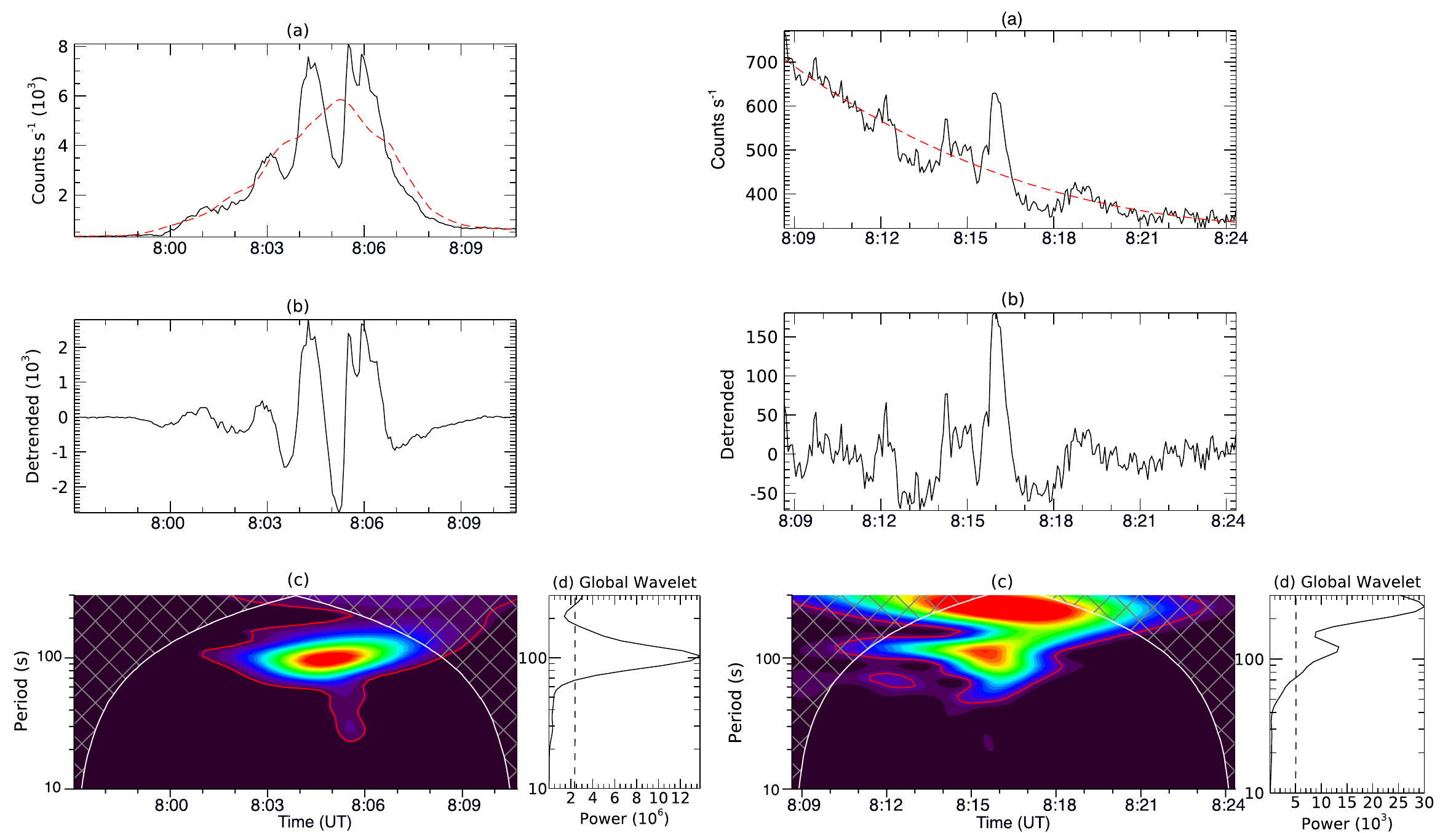}
}
\caption{{\bf X-ray QPPs associated with the ejection and merging of blobs during the first flare (C3.8).} (a-d) The wavelet analysis of the Fermi X-ray light curves (12-25 keV) in different time intervals. {\it left:} (a) X-ray light curves in the Fermi 12-25 keV channel. (b) The detrended light curve after subtracting the red trend shown in (a) from the original light curve. (c) Wavelet power spectrum of the detrended signal. Red contours outline the 95\% significance level. (d) Global wavelet power spectrum. The dashed line is the 95\% global confidence level. {\it right:} The same analysis for another interval. } 
\label{app-fig3}
\end{figure*}

\begin{figure*}
\centering{
\includegraphics[width=18cm]{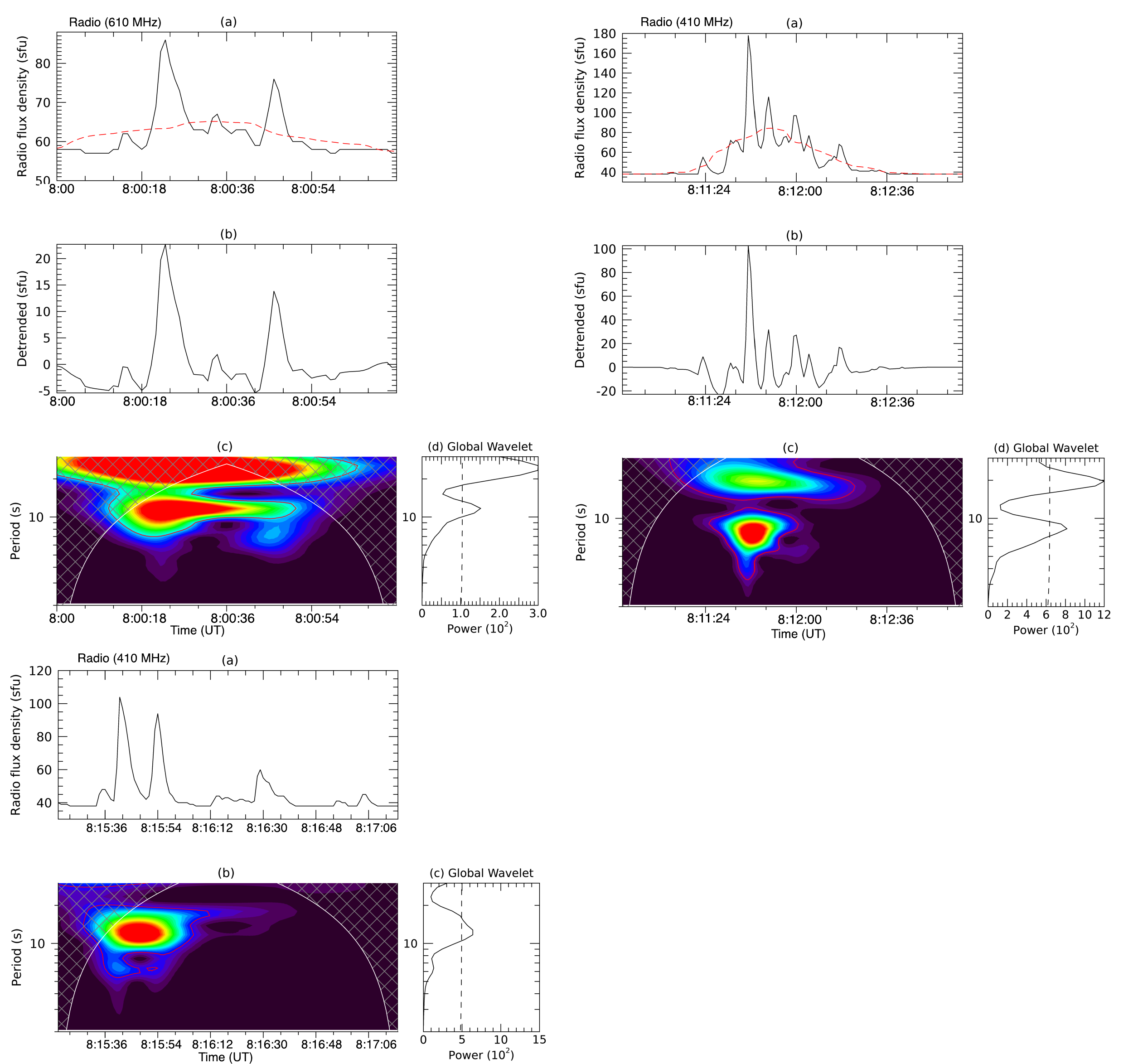}
}
\caption{{\bf Radio (decimetric) QPPs associated with the ejection and merging of blobs during the first flare (C3.8).} (a-d) The wavelet analysis of the radio flux density from Learmonth observatory in different time intervals. {\it left:} (a) Radio flux density (610 MHz) in solar flux units (1 sfu=10$^{-22}$ W m$^{-2}$ Hz$^{-1}$). (b) The detrended light curve after subtracting the red trend shown in (a) from the original light curve. (c) Wavelet power spectrum of the detrended signal. Red contours outline the 95\% significance level. (d) Global wavelet power spectrum. The dashed line is the 95\% global confidence level. {\it right and bottom left:} The same analysis for two other intervals.} 
\label{app-fig4}
\end{figure*}

\begin{figure*}
\centering{
\includegraphics[width=18cm]{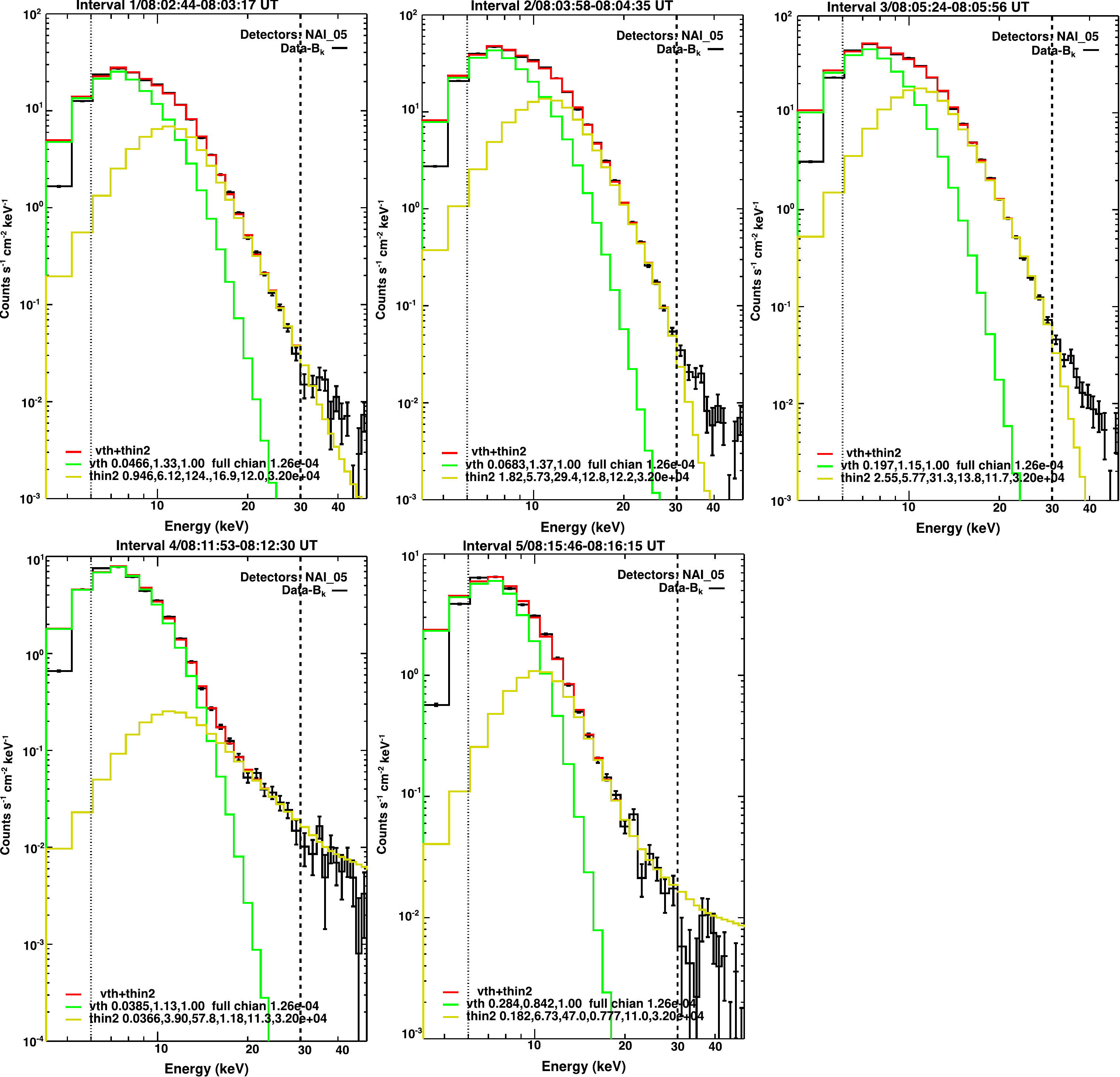}
}
\caption{{\bf Fermi GBM X-ray spectra at five different intervals during the first flare (C3.8).} The background-subtracted spectra are fitted with variable thermal (Vth) and thin-target bremsstrahlung (thin2) components. The two vertical dashed lines indicate the energy range used to fit the spectrum. } 
\label{app-fig5}
\end{figure*}

\begin{figure*}
\centering{
\includegraphics[width=12cm]{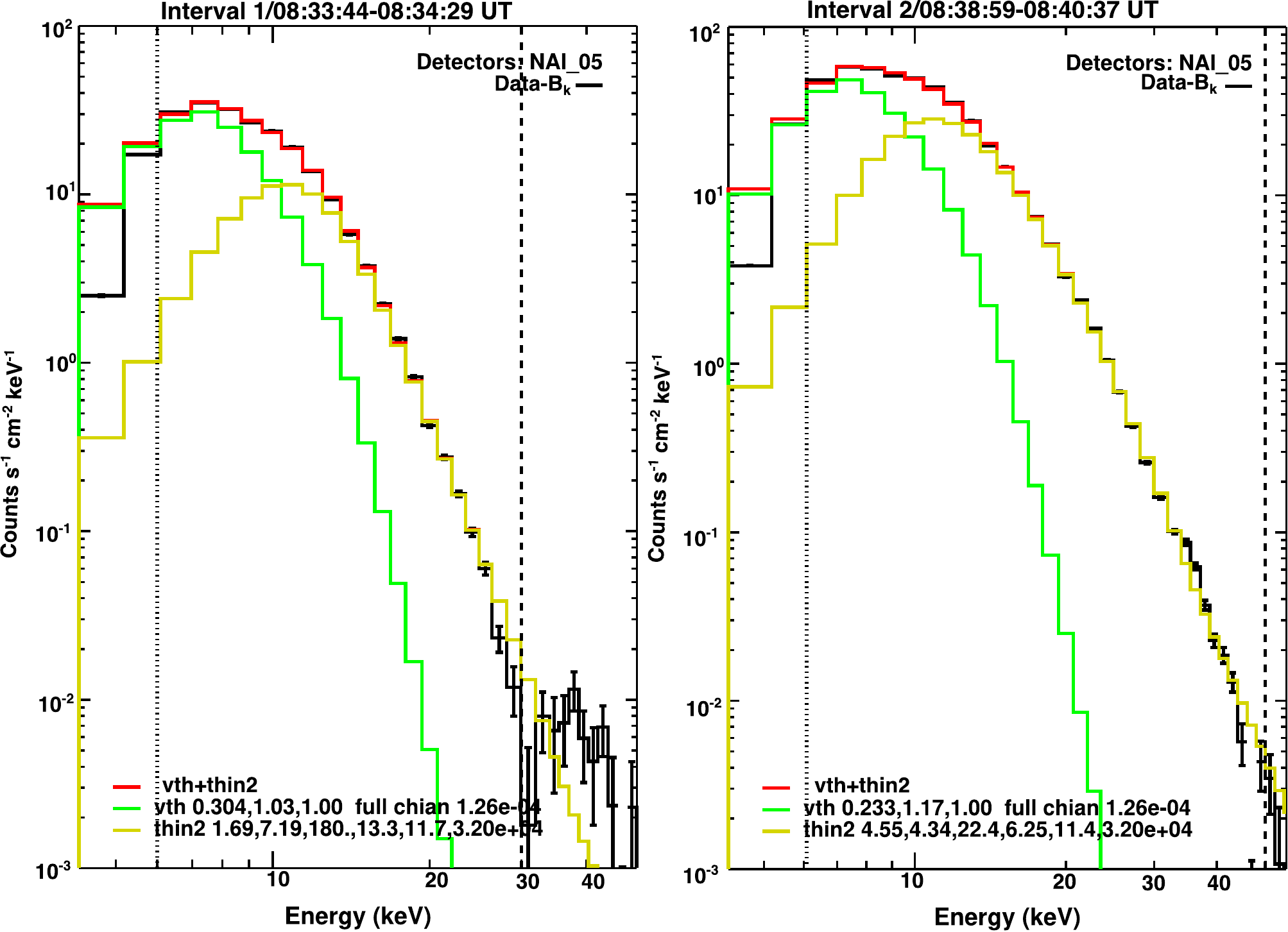}
}
\caption{{\bf Fermi GBM X-ray spectra at two different intervals during the second flare (M1.1).} The background-subtracted spectra are fitted with variable thermal (Vth) and thin-target bremsstrahlung (thin2) components. The two vertical dashed lines indicate the energy range used to fit the spectrum.} 
\label{app-fig6}
\end{figure*}

This appendix contains additional figures to support the results described above. Figures \ref{app-fig1} and \ref{app-fig2} display the high-resolution (zoomed view) dynamic radio spectra (e-Callisto) and X-ray flux in 12-25 keV (Fermi) during both flares. Figures \ref{app-fig3} and \ref{app-fig4} depict the results of our wavelet analysis for the QPPs detected in X-ray (Fermi) and radio (RSTN) wavelengths. Figures \ref{app-fig5} and \ref{app-fig6} display the background-subtracted spectra at different intervals during both flares, fitted in 6-30 keV using the OSPEX (Object Spectral Executive: \citealt{tolbert2020}) package in SSWIDL. The results of spectral fitting (temperature T, emission measure EM, index of electron flux distribution ($\delta$)) are summarized in table 1.

\clearpage

\clearpage
\clearpage
{\small
\begin{longtable*}{c c c c c}
\caption{Spectral fitting results} \\
\hline \\
\label{tab1}
Flares   &Intervals    & T              & EM                         &Index     \\
           &           & (MK)          &(10$^{49}$ cm$^{-3}$) &                  \\
         \hline
 C3.8  &1, 2, 3     &15.4, 15.8, 13.3 &0.046, 0.068, 0.019   &6.1, 5.7, 5.7  \\ 
        &4,5         &13.1, 9.7       &0.038,0.28           &3.9, 6.7       \\
 M1.1  &1, 2         &12, 13.5       &0.30,0.23         &7.1, 4.3     \\  
                                 
\hline
\end{longtable*}
\
}

\section{DEM analysis}

\begin{figure*}
\centering{
\includegraphics[width=15cm]{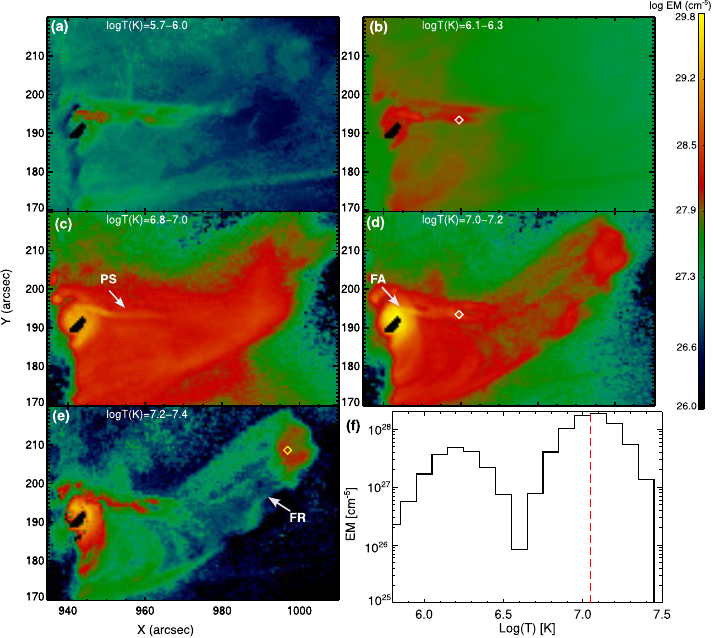}
}
\caption{{\bf DEM maps of the plasma sheet and surroundings at various temperatures} derived using near-simultaneous six-channel AIA images at 08:04:11 UT during the first flare (C3.8). The arrows indicate the bright plasma sheet (PS), flare arcade (FA), and flux rope (FR). The color coding indicates the total EM within the log(T) range marked in each panel. (f) EM profile of the plasma sheet (marked by a white diamond in panel (b)). The vertical dashed lines indicate the EM peak at log(T(K))=7.0.} 
\label{app-fig7}
\end{figure*}

\begin{figure*}
\centering{
\includegraphics[width=15cm]{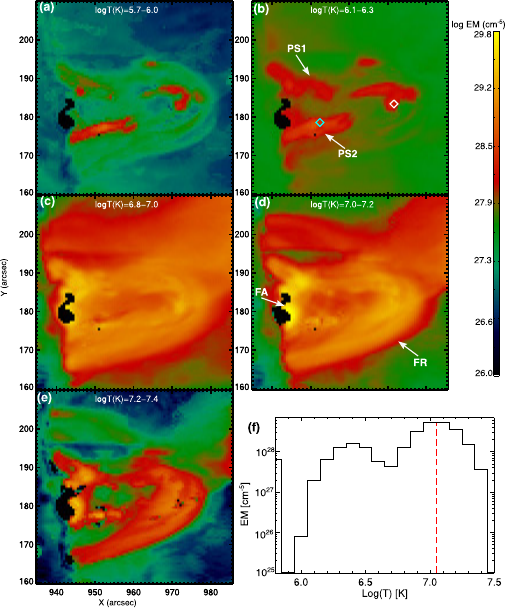}
}
\caption{{\bf DEM maps of the plasma sheets and surroundings at various temperatures} derived using near-simultaneous six-channel AIA images at 08:38:56 UT during the second flare (M1.1). The arrows indicate the bright plasma sheets (PS1, PS2), flare arcade (FA), and flux rope (FR). The color coding indicates the total EM within the log(T) range marked in each panel. (f) EM profile of the plasma sheet (marked by a white diamond in panel (b)). The vertical dashed lines indicate the EM peak at log(T(K))=7.0. } 
\label{app-fig8}
\end{figure*}

We performed a differential emission measure (DEM) analysis \citep{cheung2015} of the regions of interest using nearly cotemporal AIA images in six EUV channels (171, 131, 94, 335, 193, 211~{\AA}) at 08:04:11 and 08:38:56 UT, when the bright plasma sheets with blobs were best visible during both flares. The DEM code utilizes a log(T) grid ranging from log(T(K))=5.7 to log(T(K))=7.7, with intervals of log(T(K)) = 0.1. The emission measures in different temperature bins are presented in Figures \ref{app-fig7} and \ref{app-fig8}. The DEM distribution of blobs in the plasma sheet shows peaks at log(T(K))=6.2, 7.0 (Figure~\ref{app-fig7}(f)) and log(T(K))=6.4, 7.0 (Figure~\ref{app-fig8}) during the first and second flares, respectively. The estimated total EM, calculated by integrating the DEM distribution over the entire Gaussian temperature range, are 0.884$\times$10$^{29}$ cm$^{-5}$, 2.1$\times$10$^{29}$ cm$^{-5}$ (for blobs in PS2, cyan diamond), 2.9$\times$10$^{29}$ cm$^{-5}$ (upward moving blob below the rising flux rope, white diamond). Assuming that the depth of each blob in the plasma sheets along the line of sight is roughly equivalent to its width, w$\approx$2$\arcsec$, the electron number density for the blobs in plasma sheets
is $n = \sqrt{{EM}/{w}}$ 
= 2.4$\times$10$^{10}$ cm$^{-3}$, 3.7$\times$10$^{10}$ cm$^{-3}$, 4.3$\times$10$^{10}$ cm$^{-3}$ (assuming filling factor=1). The estimated total EM for the flux rope is 0.26$\times$10$^{29}$ cm$^{-5}$ and a width of $\approx$10 arcsecs. Similarly, the estimated density of the flux rope is about 5.8$\times$10$^{9}$ cm$^{-3}$ (Figure \ref{app-fig7}(e)).
The bright plasma sheets and blobs (PS, PS1/PS2 marked by arrows) display temperatures between 10-25 MK ($\log T \mathrm{(K)}=7.0$--$7.4$) (Figures~\ref{app-fig7} and Figures~\ref{app-fig8} (d,e)). The flux rope (FR) and hot arcade loops (FA) are most clearly observed at above 10 MK. The AIA images and DEM maps also show the blobs in cool/warm temperatures, suggesting the existence of multithermal plasma.
These values are consistent with previous estimates of densities in plasma sheets and blobs \citep{kumar2013,warren2018,kumar2024}. 

\clearpage

\section{Supplementary Materials}
This section contains supplementary movies to support the results. All high-quality supplementary movies are available in the Zenodo repository at doi:\href{https://doi.org/10.5281/zenodo.14286164}{10.5281/zenodo.14286164}. \\
{\bf Movie S1}: An animation of the AIA 171, 211, and 131 {\AA} images (Figure \ref{fig2}) covers the first flare (C3.8). The top panels show a large field of view, while the lower panels display a zoomed-in view of the area outlined by the red box in the top panels. The animation runs from 07:51:37 UT to 08:26:01 UT. Its real-time duration is 5.7 s. \\
{\bf Movie S2}: The zoomed-in view of AIA 171 and 211 {\AA} animation (left panels, Figure \ref{fig4}(a1-d2)) depicts the plasma sheet and the ejection of multiple plasmoids during the first flare (C3.8). The right panels display the ORFESS radio dynamic spectrum (decimetric bursts, Figure \ref{fig3}(d)) and Fermi GBM X-ray flux (6-12, 12-25, and 25-50 keV) during the first flare (Figure \ref{fig3}(c)). The animation runs from 07:51:37 UT to 08:26:01 UT. Its real-time duration is 5.7 s. \\
{\bf Movie S3}: The AIA 171, 211, and 131 {\AA} animation (left panels, Figures \ref{fig4} and \ref{fig5}) and the ORFESS radio dynamic spectrum (Figure \ref{fig3}(d)), Fermi GBM X-ray flux (Figure \ref{fig3}(c)), and RHESSI images in the 6-12 keV channel (bottom rows of Figures \ref{fig4} and \ref{fig5}) during the first flare (C3.8). The animation runs from 08:10 UT to 08:20 UT. Its real-time duration is 2.2 s.  \\
{\bf Movie S4}: An animation of the AIA 171, 211, and 131 {\AA} images (Figure \ref{fig6}(b1-d1)) covers the second flare (M1.1). The top panels show a large field of view, while the lower panels display a zoomed-in view of the area outlined by the red box in the top panels. The animation runs from 08:26:13 UT to 08:56:37 UT. Its real-time duration is 5.1 s.\\
{\bf Movie S5}: The AIA 171, 211, and 131 {\AA} animation (Figure \ref{fig6}(b1-d4)) is shown in the left panels, while the right panels display the Fermi GBM X-ray flux (6-12 keV, Figure \ref{fig3}(c)), GOES soft X-ray flux derivative (1-8~{\AA}) (Figure \ref{fig1}(d)), and RHESSI images in the 6-12 keV channel during the initiation of the second flare (M1.1). The animation runs from 08:26:25 UT to 08:33:37 UT. Its real-time duration is 3.8 s.\\
{\bf Movie S6}: The first part of the animation shows a TD intensity plot along P1Q1 (Figures \ref{fig7}(b4) and \ref{fig8}(a,b)) using AIA 211~{\AA} images and the associated e-Callisto radio dynamic spectrum (decimetric bursts) during the second flare (M1.1). The second part of the animation covers TD intensity plots along P2Q2 and P4Q4 using AIA 211~{\AA} images (Figures \ref{fig7}(b5,b6) and \ref{fig8}(c,e)). The third part of the animation shows a TD intensity plot along slice P3Q3 using AIA 131~{\AA} images (Figures \ref{fig7}(c6) and \ref{fig8}(d)). The animation runs from 08:20 UT to 08:50 UT. Its real-time duration is 39 s.\\
{\bf Movie S7}: The animations (a,b) show density (intensity and running difference) images from a 3D MHD simulation with ARMS (Figure \ref{fig9}), highlighting the formation and ejection of multiple plasmoids moving bidirectionally in the flare current sheet beneath an erupting flux rope. The animations run from t=12800-14195 s and real time duration is about 22 s. \\
\end{document}